\documentclass[12pt,a4paper]{article}
\usepackage[margin=25mm]{geometry}                % See geometry.pdf to learn the layout options. There are lots.
\usepackage{graphicx}
\usepackage{amsmath}
\usepackage{amssymb}
\usepackage{epstopdf}
\usepackage{hyperref}
\usepackage{caption}
\usepackage[percent]{overpic}
\usepackage{longtable}    % so it can break across pages 
\usepackage{pdflscape,afterpage}  % to set it in landscape

\title{Parametric study of {\em E. coli} incidence with reference to the New Zealand freshwater standards and the Manawat\={u}--Whanganui region}
\author{Stephen R Marsland$^1$, Robert I McLachlan$^2$, and Christopher Tuffley$^2$}
\date{%
$^1$ School of Mathematics and Statistics, Victoria University of Wellington, New Zealand\\
$^2$ School of Fundamental Sciences, Massey University, New Zealand\\
}
\begin{document}
\maketitle
%\section{}
%\subsection{}

\begin{abstract}
The New Zealand National Policy Statement for Freshwater Management 2020 sets several targets for freshwater quality, six of which are measurements of rivers; others relate to lakes. Each regional council is required to monitor freshwater quality and to respond as prescribed in order to meet the targets. One target of particular public interest  is based on four criteria determined from recent  {\em E. coli} readings, and concerns the health risk of swimming in a river. However, the inherent variability of the data makes it difficult to  determine the water quality state and trend reliably, particularly using traditional methods based on percentiles. Therefore, in this study we return to the parametric lognormal model of {\em E. coli} distribution, from which the official criteria were developed. We interpret the classification system in terms of the parametric model and show that the parametric model can reduce uncertainty and can incorporate more useful information, especially from very high {\em E. coli} readings, and is suitable for censored data. We apply the parametric model for state and trend to 135 sites in the Manawat\={u}--Whanganui region.
\end{abstract}

\section{Introduction}

In August 2017, the New Zealand Government approved amendments to the National Policy Statement (NPS) for Freshwater Management---the `Clean Water' package. These amendments, in particular the criteria for a river to be deemed `swimmable', had been the subject of extensive background research and were also of great public interest. Their adoption followed a period of some decades of refinements. Historical, public health, international comparison,  and  statistical and freshwater science issues related to the NPS are discussed by McBride \cite{mcbride}. The subject continues to be of national importance, as each regional council in New Zealand seeks to implement the NPS and to manage its freshwater monitoring and improvement programme. A new NPS, finalising the criteria, came into force on 3 September 2020 \cite{cabinet}.

The swimmability criteria place rivers into 5 categories (A--E, also called Blue, Green, Yellow, Orange, and Red, of which A, B, and C are deemed `swimmable'), based on four {\em E. coli} criteria, shown in Table~\ref{tab:cri}\footnote{The cabinet paper clarifies that ``all four measures must be met and should be determined using a minimum of 60 samples over a maximum of
five years collected on a regular basis regardless of weather and flow
conditions. The only exception to this is if there is insufficient monitoring data to
establish the 95th percentile, in which case that measure/statistic does not apply.''}

Two of the four criteria are based on percentiles (P50 and P95, i.e., the 50th and 95th percentiles of the {\em E. coli} counts), and two are based on `percentage exceedances': G260 and G540 refer to the percentage of samples that should not exceed bacteria counts of 260 and 540, respectively. There is a further classification for sites that are likely to be used for swimming, based solely on 95\% thresholds. In practice this classifies  B, C, D, and E sites as Poor, and subdivides A sites into Excellent, Good, and Fair. We call the standard method based on P50, P95, G260, and G540 the {\em percentile classification}. 

Regional councils are responsible for measuring and reporting against all targets. The data is generally collected monthly, from a set of pre-defined locations across the river systems of the region. These locations may be chosen for a variety of reasons, including upstream and downstream points close to potential polluters, convenience of sampling, and proximity to human settlements. New measurement points may be added if pollution is detected in a particular place. 

{\em Escherichia coli (E. coli)} are bacteria found in the digestive tract of warm-blooded animals, some of which can cause sickness in humans. {\em E. coli} readings in freshwater, usually given as the number of bacteria present in a 100 mL sample, are highly variable. {\em E. coli}  generally gets into rivers from human sewage outlets or from farm run-off. It then flows downstream. If its presence in the river is because of a contamination episode such as heavy rain or other adverse weather event, it will be present in the river system for only a few days, but if it comes from a persistent polluter it will be present continuously. Usually, monthly readings are essentially uncorrelated in time. A widely-used approach models {\em E. coli} readings as independent and lognormally-distributed random variables. For example, S\o rensen {\em et al.} \cite{so-ja-sp} use the lognormal parametric model to develop a sampling protocol for swimmability. 

Moreover, {\em E. coli} concentrations are only a proxy for freshwater quality. The guidelines were developed by modelling the risk of {\em Campylobacter} infection {\em using} the lognormal parametric model of {\em E. coli} (see, e.g., \cite[p. 48]{cabinet}.) The thresholds were determined from a combination of this risk, historical precedent, and international comparisons. Percentile criteria (and categories) were adopted for easier comprehension by the public and to be clear and simple for councils to measure and report. This process of turning a multidimensional state into discrete ordinal categories (sometimes called {\em dichotomization} or {\em ordinalization}) necessarily loses information; the ordinal data should almost never be used for subsequent statistical analysis \cite{fedorov}. In addition it should be remembered that the goal of this and any other clean water strategy is not the specific targets (e.g., a fraction of rivers by length to be `A' quality by a certain date), but an improvement in the underlying multidimensional state. A further complication is that the water quality criteria are used for several different purposes, namely (i) determining the state of a particular site; (ii) determining the trend in state of a particular site; (iii) determining a broad measure of the state or trend of all measured sites; (iv) predicting the state or trend of all rivers in the region.

\begin{table}
\begin{center}
\begin{tabular}{ l  cccc }
\hline\hline
Category & G540 & P50 & P95 & G260 \\
& samples above  540 & median & 95th percentile & samples above 260 \\
%Excellent & & & $\le 130 & \\
%Good & & & $\le 260 & \\
%Fair & & & $\le 540 & \\
%Poor & & & $>540 $\\
\hline
A | Blue & $<5\%$ & $\le$ 130 & $\le$ 540 & $<20\%$ \\
B | Green & 5--10\% & $\le 130$ & $\le 1000$ & 20--34\%\\
C | Yellow & 10--20\% & $\le 130$ & $\le 1200$ & 20--34\% \\
D | Orange & 20--30\% & $>130$ & $>1200$ & $>34\%$ \\
E | Red & $>30\%$ & $>260 $ & $>1200$ & $>50\%$ \\
\hline\hline
\end{tabular}
\end{center}
\caption{\label{tab:cri}{\em E. coli}  criteria in the National Policy Statement for freshwater. Categories A, B, and C meet the national bottom line, informally known as `swimmable'. The NPS states that `Attribute state must be determined by satisfying all numeric attribute states'. (Presumably it is intended that  a site with, for example, G540 $<$5\%, median $<$130, and G260 $=$22\% falls into category B.)}
\end{table}

The Manawat\={u}--Whanganui Regional Council commissioned reports into the state and trend of regional rivers, and these were used to develop swimmability targets for 2030 and 2040 \cite{sn}. These studies examined each {\em E. coli} threshold separately, and the importance of sampling errors figure prominently. Notably, the 95th percentile criterion (P95) was discarded because it is estimated with lower precision than the other three criteria P50, G260, and G540. Trends in G260 and G540 were estimated from the trends in their annual values, which lowers their precision.

In this paper we compare the parametric lognormal model to the percentile classification and show that the parametric model can reduce uncertainty and can incorporate more useful information, especially from very high {\em E. coli} readings, and is suitable for censored data. We apply the parametric model for state and trend to 135 sites in the Manawat\={u}--Whanganui region and make recommendations for points that are worth further consideration in the analysis of swimmability.

\section{State determination in the parametric model }

\subsection{Interpretation of the NPS criteria in the parametric model}

For state estimation, we consider  a river in a steady state whose {\em E. coli} measurements at times $t_i$ are independent random variables drawn from a lognormal distribution, i.e.: $$L_i := \log(\hbox{\em E. coli}(t_i))\in N(\mu,\sigma).$$
Here $\log$ is the natural logarithm. 

The water quality criteria have a simple interpretation in terms of this parametric model. Each individual criterion corresponds to a half-plane in the $(\mu,\sigma)$ parameter space, with each category A--E being an intersection of such half-planes, which corresponds to a polygon in $(\mu,\sigma)$ parameter space.

For example, the criterion that the median be less than or equal to 130 becomes the half-plane $\mu\le \log(130)\approx4.868$,
and the criterion that 95th percentile is less than or equal to 1200 becomes the half-plane $\mu + z_{0.95}\sigma \le \log(1200)\approx 7.0890$. Here $z_p$ is the $z$-value corresponding to the quantile $p$ for the normal distribution, i.e., the inverse of the cumulative density function. From Table~\ref{tab:cri}, the relevant values are $z_{0.5}=0$, $z_{0.66}=0.412$, $z_{0.7}=0.524$, $z_{0.8}=0.842$, $z_{0.9}=1.282$, and $z_{0.95}=1.645$. 

\begin{figure}
\small
\begin{center}
\begin{overpic}[width=5cm]{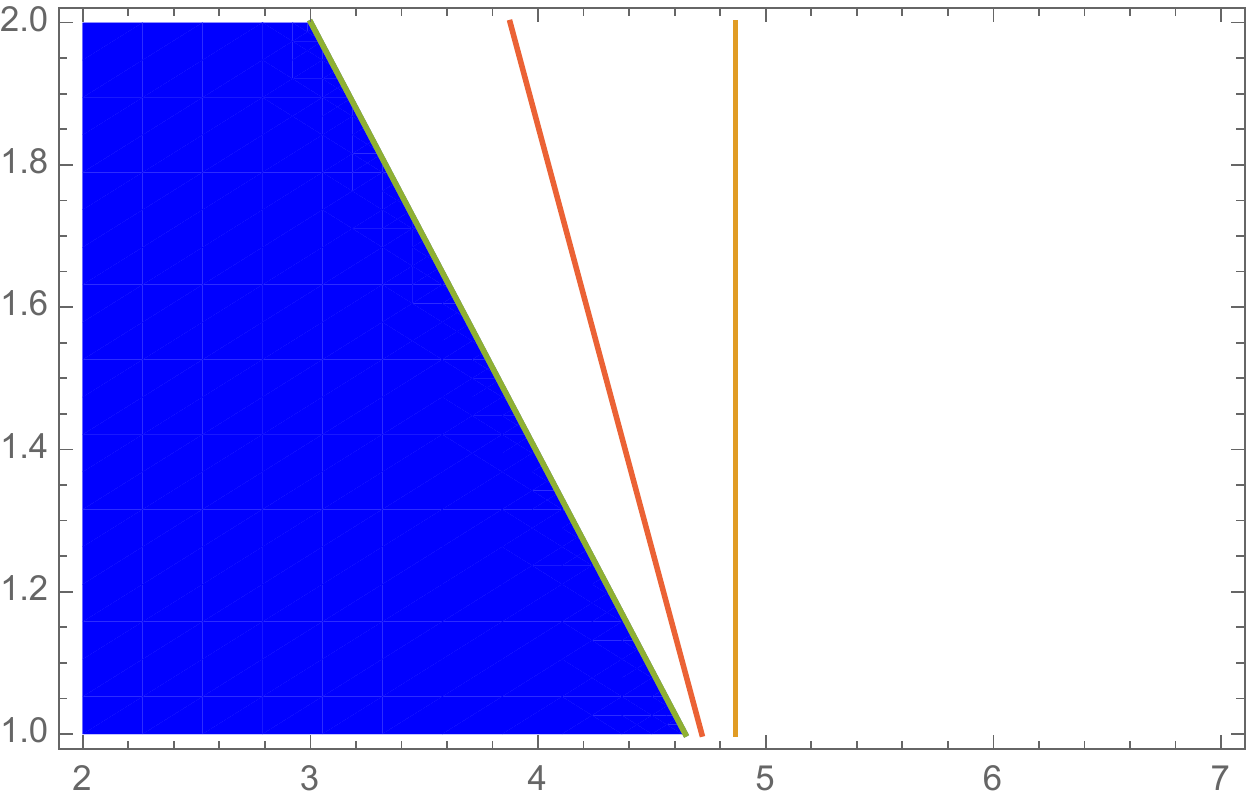}
\put (85,0){$\mu$}
\put (-5,55){$\sigma$}
\put (20,68) {Cutoffs for category A}
\end{overpic}
\hspace{8mm}
\begin{overpic}[width=5cm]{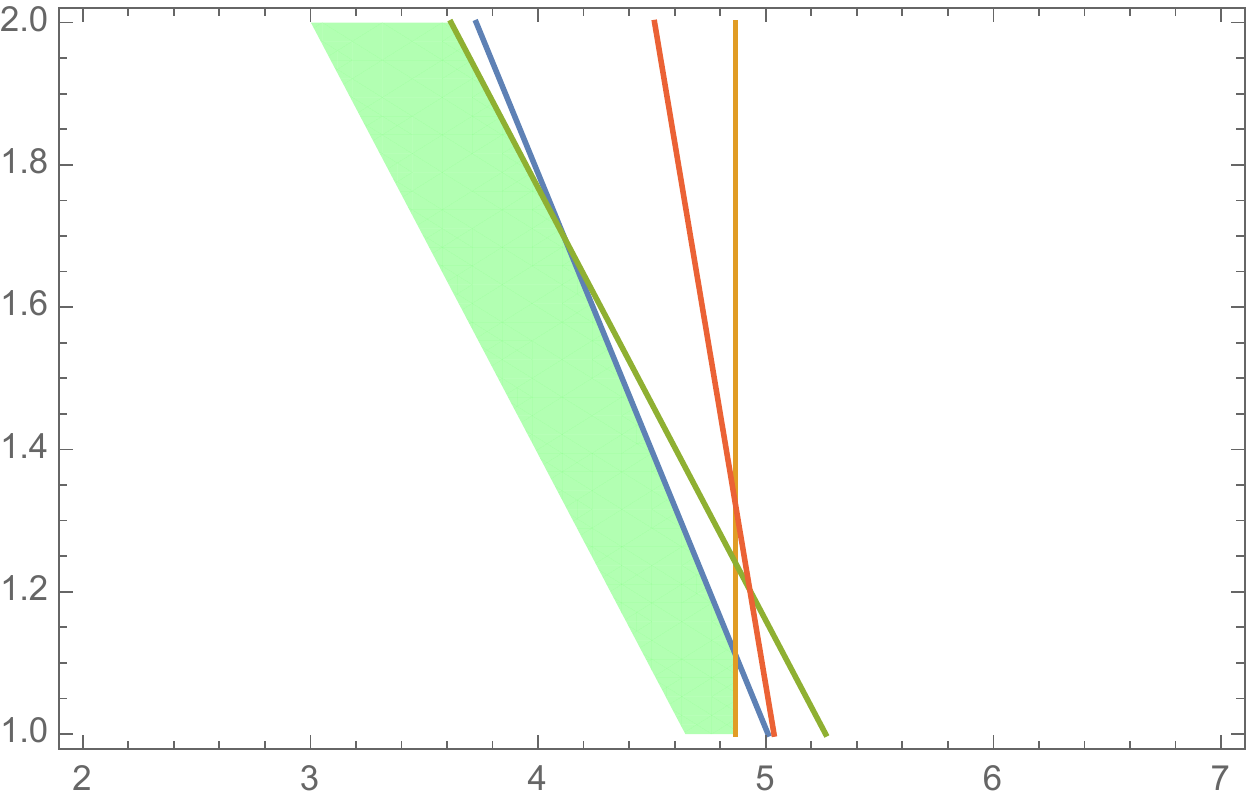}
\put (85,0){$\mu$}
\put (-5,55){$\sigma$}
\put (20,68) {Cutoffs for category B}
\end{overpic}\\
\vspace{6mm}
\begin{overpic}[width=5cm]{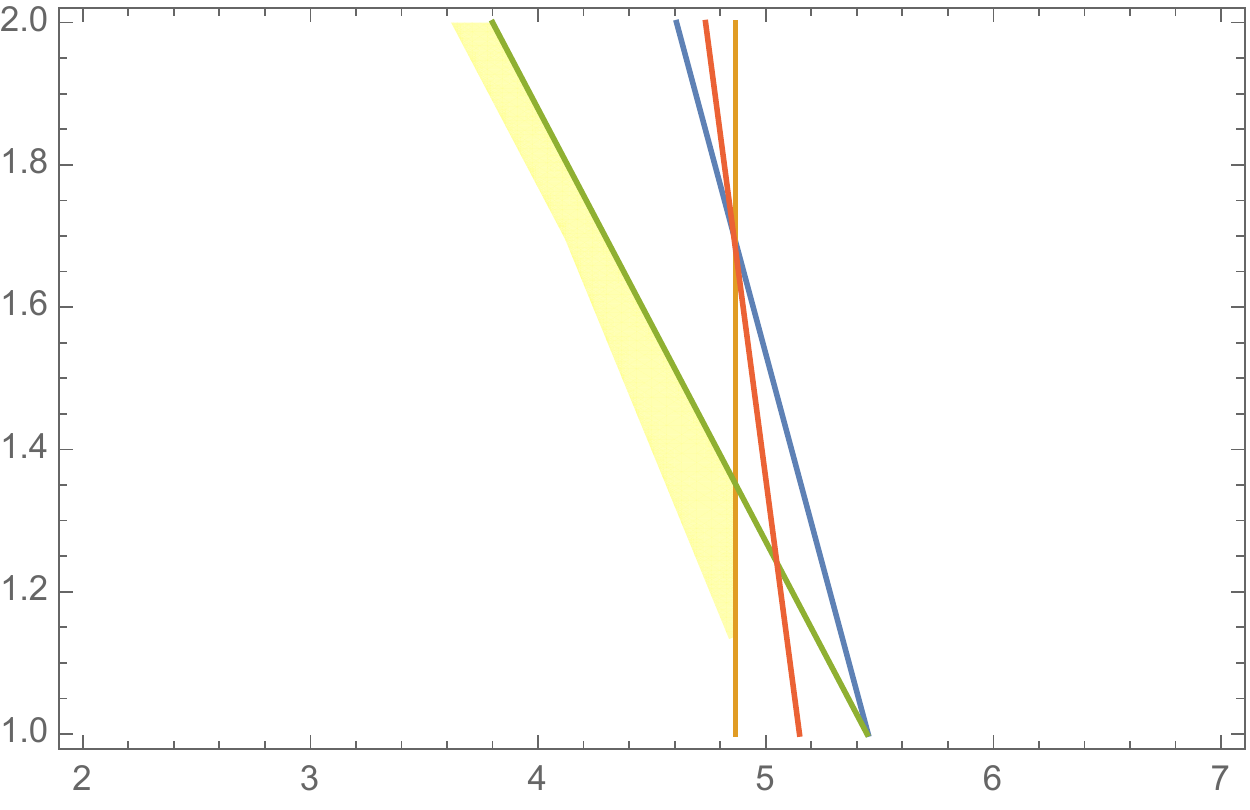}
\put (85,0){$\mu$}
\put (-5,55){$\sigma$}
\put (20,68) {Cutoffs for category C}
\end{overpic}
\hspace{8mm}
\begin{overpic}[width=5cm]{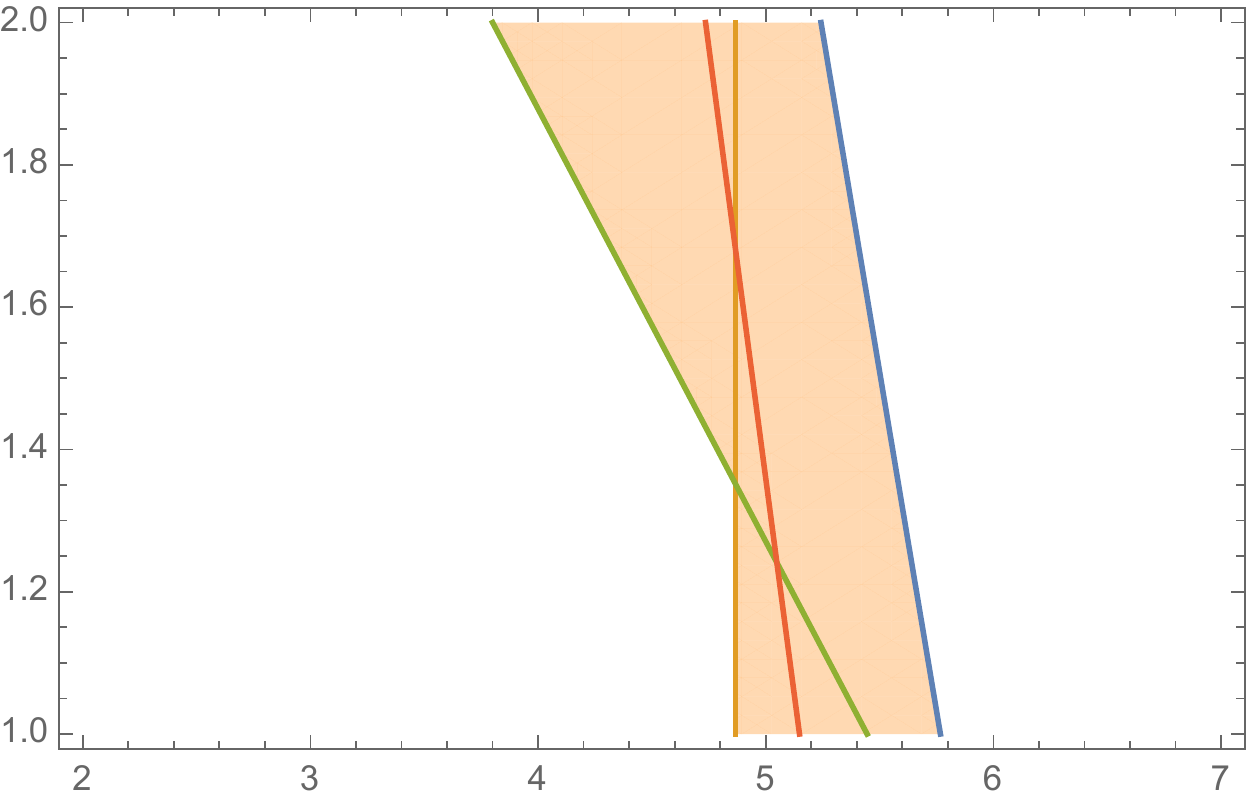}
\put (85,0){$\mu$}
\put (-5,55){$\sigma$}
\put (20,68) {Cutoffs for category D}
\end{overpic}
\vspace{6mm}

\begin{overpic}[width=5cm]{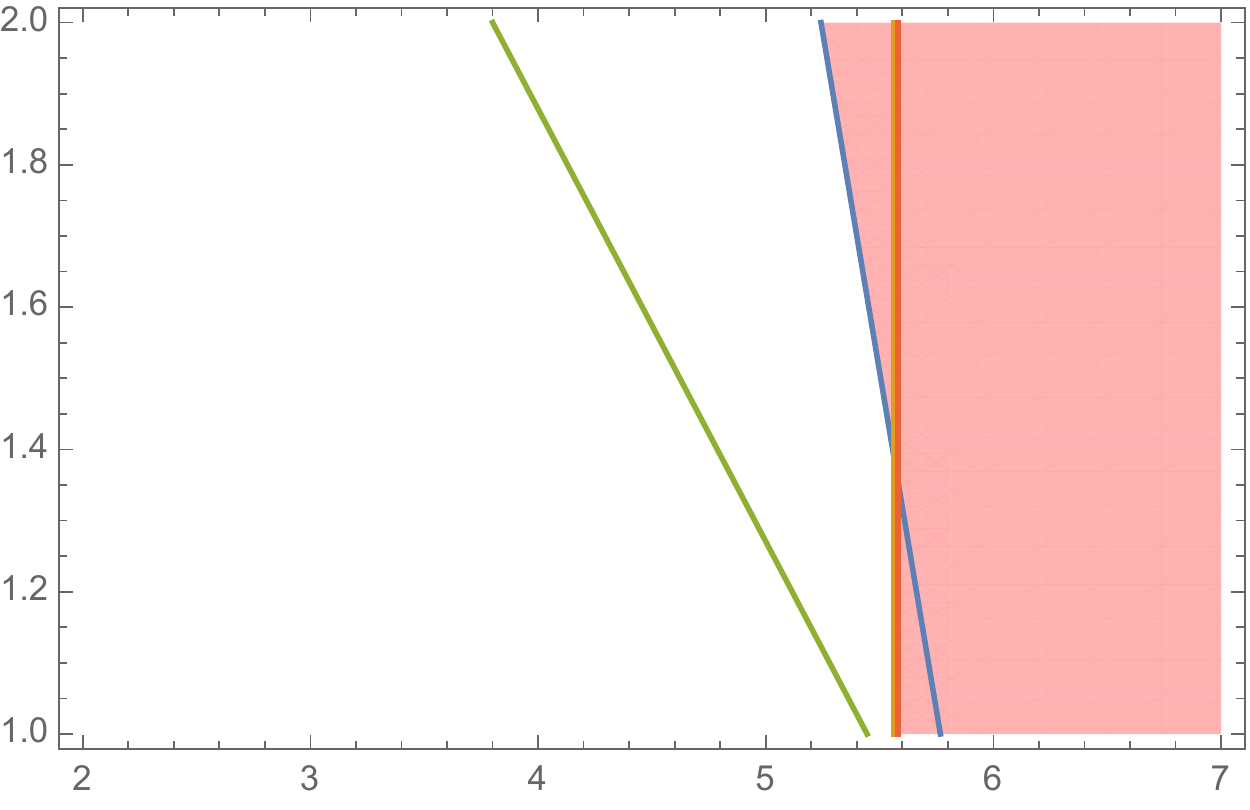}
\put (85,0){$\mu$}
\put (-5,55){$\sigma$}
\put (20,68) {Cutoffs for category E}
\end{overpic}
\hspace{5mm}
\includegraphics[width=1.4cm]{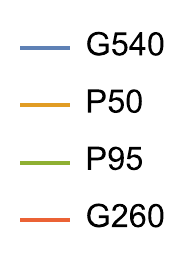}
\hfill \\
\end{center}
\caption{\label{fig:ecri}
In the parametric model, each of the four swimmability criteria for each of the five categories corresponds
to a half-plane in the $(\mu,\sigma)$ parameter space. For parameter values observed in practice, many
of these criteria are inactive. Except for the D--E boundary, overall the P50 and P95 criteria are the most important.
}
\end{figure}

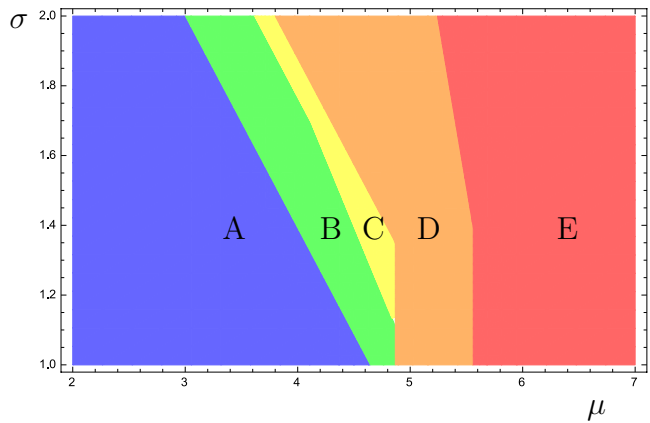
\begin{figure}
\begin{center}
\begin{overpic}[width=8cm]{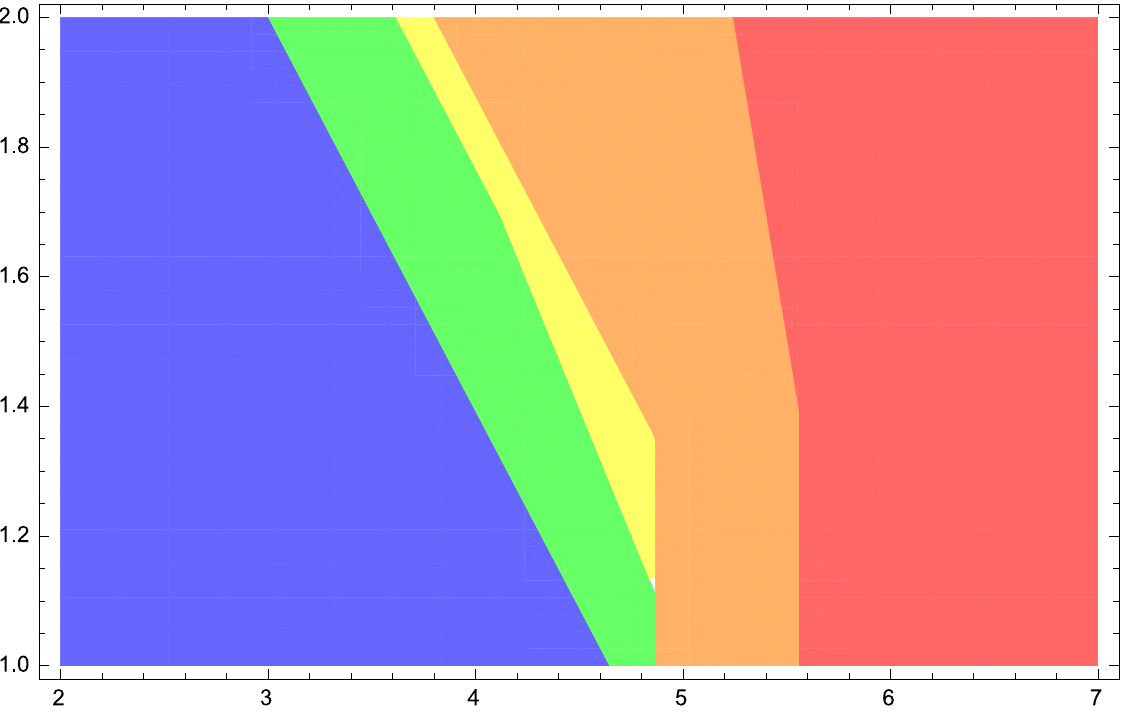}
\put(90,-4){$\mu$}
\put (-5,60){$\sigma$}
\put (30,25) A
\put (46,25) B
\put (53,25) C
\put (62,25) D
\put (85,25) E
\end{overpic}
\end{center}
\caption{\label{fig:ecriall}
The five swimmability categories visualized in $(\mu,\sigma)$ parameter space in the parametric model.
}
\end{figure}

The individual criteria are shown in terms of the parametric model in Figure~\ref{fig:ecri}, and the resulting boundaries of these polygons for swimmability categories A--E  in Figure~\ref{fig:ecriall}. We summarize them as follows:
\begin{itemize}
\item[A--B] For the A--B boundary, only the P95 criterion (identical to G540 in this case) is relevant. All sites meeting P95 already greatly exceed the requirements for P50 and G260.
\item[B--C] The B--C boundary is determined for most sites by the G540 criterion. (For very clean sites P95 becomes active, and there is also a B--D boundary that is relevant for dirty sites, determined by P50, although in our data no sites came near either of these two parts of the boundary.) As G540 here is determined by the percentage of samples being above or below 10\%, this is essentially equivalent to the 90th percentile.
\item[C--D] The C--D boundary---perhaps the most important, as it determines swimmability---is determined by P95 and P50. Sites meeting these criteria already satisfy G260 and G540.
\item[D--E] The D--E boundary is determined by G540, in this case equivalent to the 70th percentile.
\end{itemize}

Thus, the four criteria are not equally important for all sites. 

\newpage
\paragraph{Remarks}
\begin{enumerate}
\item
G260 and G540, on which a lot of attention has been placed, are likely to be irrelevant for determining whether a particular site is swimmable. 
\item The criteria have the effect of making the category C region relatively small. For the sample data, far more sites are in category B than category C, which diminishes the usefulness of the categories. 
In terms of this model, it would make more sense to replace the B--C boundary at P95 = 800 instead of P95 = 1000.
\item The crucial C--D (pass--fail) boundary is mostly determined by P95. But P95 is the least reliable of the four criteria to measure, and is often dropped.
\item Dropping the 95th percentile requirement for the D--E boundary in favour of G540 requirement does make sense in this model.
\end{enumerate}

\subsection{Sampling errors in the parametric and percentile models}
When $n$ samples are drawn from a normal distribution with mean $\mu$ and standard deviation $\sigma$, the sampling
error in the mean is $\sigma/\sqrt{n}$, and the sampling error in the standard deviation is $\sigma/\sqrt{2n}$. Therefore,
the sampling error in $\mu + z_p\sigma$, the natural estimate of the $p$th percentile under the normal distribution, is
$d_p\sigma/\sqrt{n}$ where $d_p = \sqrt{1+\frac{1}{2}z_p^2}$.

There are different methods of estimating percentiles from data. In {\em E. coli} studies, this is often done by applying the Hazen method to the log of the bacteria counts \cite{hunter}, which is implemented in software provided by the Ministry for the Environment \cite{mcbride}.

\begin{figure}
\begin{center}
\includegraphics[width=5cm]{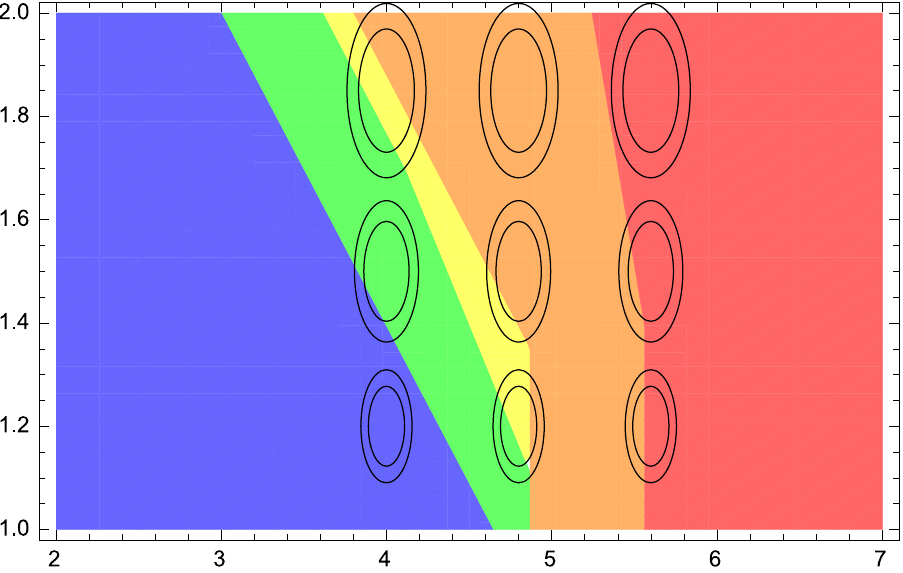}
\hspace{6mm}
\includegraphics[width=5cm]{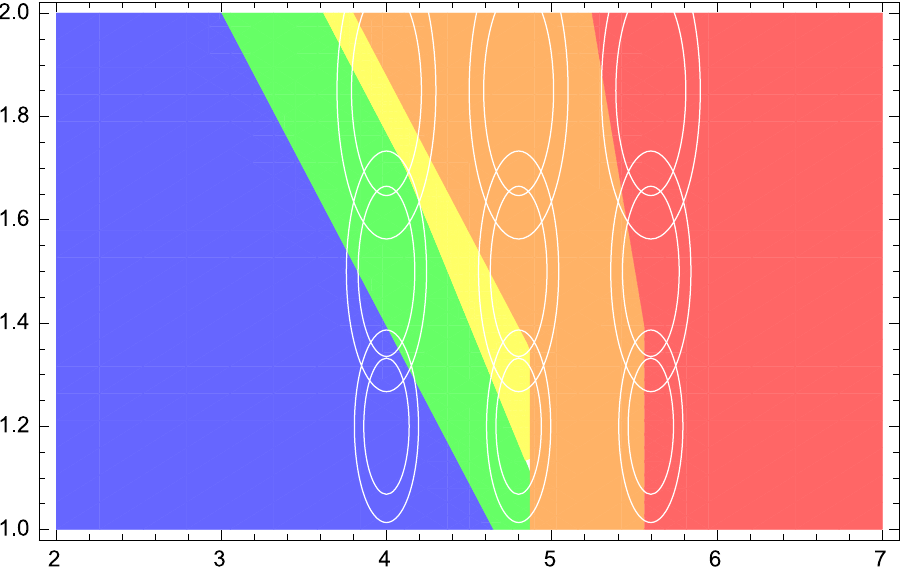}
\end{center}
\caption{\label{fig:ci}
An illustration of 67\% (1-sigma) confidence intervals when estimating the state of a single site with
either 60 (large ellipses) or 120 (small ellipses) independent samples. Left: using the estimates
of $(\mu,\sigma)$ from the parametric model. Right: Using percentiles in the parametric model leads
to larger confidence intervals and larger sampling errors.
}
\end{figure}

Sampling errors in percentiles are larger than sampling errors in $\mu+z_p\sigma$, as was pointed out in a study based on simulations \cite{hunter}. The precise statement is that
the sampling error in the $p$th percentile for a distribution with probability density function $f$ and cumulative density function $F$ is $c_p \sigma/\sqrt{n}$ where $c_p = \sqrt{p(1-p)}/f(F^{-1}(p))$ \cite{kendall}.
For example, the sampling error in the median for a normally-distributed random variable is $\sqrt{\frac{\pi}{2}}/\sqrt{n}$, a factor of 1.253 times larger
than the sample error in the mean: see Figure~\ref{fig:ci} for an illustration of the differences as they affect the swimmability categories. Comparisons for different percentiles are given in Table~\ref{tab:sampl}. 

The above values are approximations that are valid in the limit of large $n$. For small $n$, the bias of these estimators may be significant as well. The 50th percentile estimators (the median and mean, respectively) are unbiased. The Hazen percentiles are biased: for example, for normal data the 95th percentile
has bias $-0.07\sigma$ when $n=12$ and $-0.008\sigma$ when $n=60$. In contrast, the sample mean $\bar x$ and sample standard deviation $s$ easily provide the unbiased estimator $\bar x + \alpha s z_p$, where $\alpha = \sqrt{\frac{n-1}{2}}\Gamma\left(\frac{n-1}{2}\right)/ \Gamma\left(\frac{n}{2}\right)$
\cite{parrish}.

\begin{table}
\begin{center}
\begin{tabular}{c  ccc }
\hline\hline
$p$ & $c_p$ & $d_p$ & sampling factor \\
\hline
0.5 & 1.253 & 1 & 1.570 \\
0.66 & 1.293 & 1.042 & 1.540 \\
0.7 & 1.318 & 1.067 & 1.526 \\
0.8 & 1.429 & 1.163 & 1.510 \\
0.9 & 1.709 & 1.350 & 1.603 \\
0.95 & 2.113 & 1.534 & 1.897\\ \hline\hline
\end{tabular}
\caption{\label{tab:sampl} Coefficients of the standard error in the $(100p)$th percentile of data in $N(\mu,\sigma)$ in the limit of large $n$, as calculated by the percentile method (standard error $c_p\sigma/\sqrt{n}$) and from the parametric method (standard error $d_p\sigma/\sqrt{n}$), together with the ratio between the number of samples required by the percentile and parametric methods for the same sample error (so for the 95th percentile, the percentile method needs $(2.113/1.534)^2 = 1.897$ times as many samples to achieve the same sampling errors).}
\end{center}
\end{table}

\begin{figure}
\begin{center}
\begin{overpic}[width=6cm]{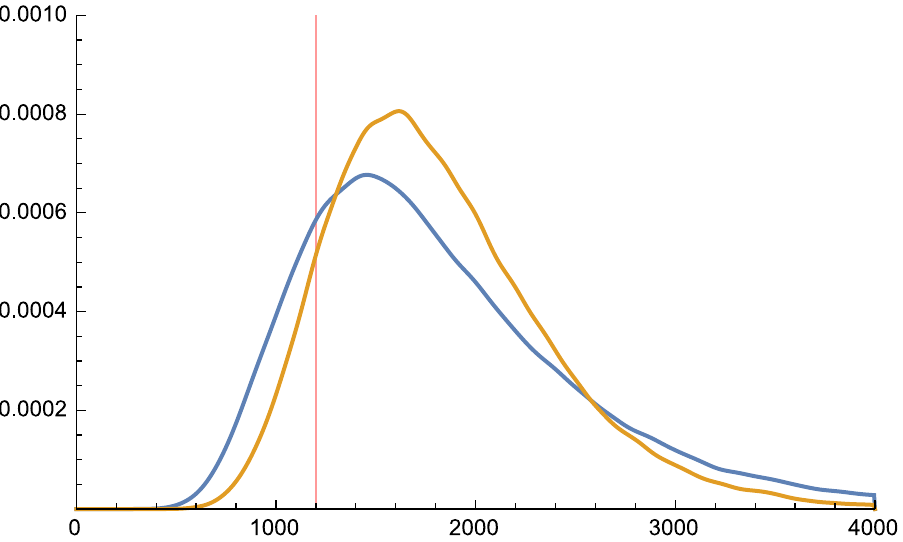}
\put (85,-4) {\small \em E. coli}
\put (-8,52) {\small PDF}
\end{overpic}
\end{center}
\caption{\label{fig:samples}
The distribution of estimates of the {\em E. coli} 95th percentile obtained by the parametric (orange) and  percentile (blue) methods for synthetic data corresponding to a 
a D-category site with median {\em E. coli} count of 150 and 95th percentile count of 1750. 
With 60 independent samples, the two methods falsely report P95$<$1200 18\% (percentile) and 10\% (parametric) of the time.
}
\end{figure}

An example is shown in Figure~\ref{fig:samples} for synthetic data generated for a site with mean {\em E. coli} count of 150 and P95 count of 1750.
With 60 independent samples, the percentile method falsely reports P95 $<$ 1200 18\% of the time;
in contrast, the parametric method falsely reports P95 $<$ 1200 only 10\% of the time.

\subsection{{\em E. coli} data for the  Manawat\={u}--Whanganui region}
Data were obtained from LAWA \cite{lawa} for 135 sites in Manawat\={u}--Whanganui. The 18,567 measurements ranged from January 2005 to December  2019. The number of observations per site ranged from 15 (there were 4 sites with fewer than 60 data points) to 189, with a median of 145 (i.e., 12 years of monthly sampling data). Observations are generally monthly with some missing observations.

\begin{figure}
\begin{center}
\includegraphics[width=12cm]{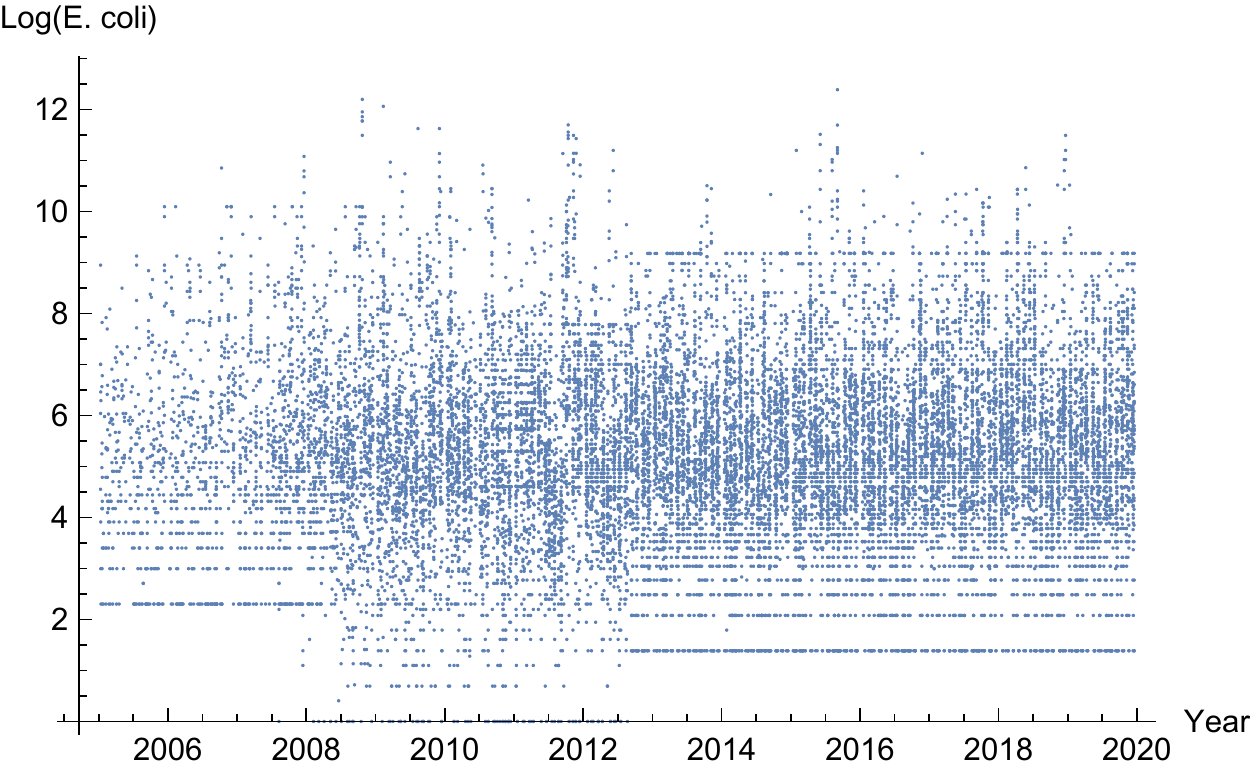}
\caption{\label{fig:alldata} The 18,567 measurements of {\em E. coli} from 135 sites. }
\end{center}
\end{figure}

A plot of all the data points for all 135 sites is shown in Figure~\ref{fig:alldata}. Changes in the measurement (or reporting) system over time can be seen, particularly with regard to precision and censoring (that is, reporting only a range in which the value lies).  The reported precision varies with both time and value. Prior to 2008, mid-range values (near 400) were rounded to the nearest 5 units (i.e., a precision of 1\%); values less than 60 were rounded to the nearest 10 units; large values were reported as 9,800, 13,000, 14,100, 17,300, 19,900, 24,200, or 51,700. Some readings were censored, i.e., reported as `$<$10' or `$>$24,200'.

From 2008 to mid-2012, there was no rounding, and the only censoring involved values reported as `$<$1' (presumably because no bacteria were observed in the 100 mL sample). However, from mid-2012, values up to 2300 were reported to 2 significant figures, and higher values with gradually decreasing precision. High values for most sites were reported as 6,200, 6,900, 7,900, and 9,700. Low values were censored and reported as `$<$4'. Most high values were censored and reported as `$>$9,700', although some sites continued to report higher values (up to 120,000, and `$>$240,000').

Rounding is presumably related in some way to the underlying accuracy of the measurements; analytic errors in {\em E. coli} measurement are typically around 20\% \cite{angelescu}. However, the rounding itself adds errors of 10--20\%, especially  for low and high values that are reported with less precision (consider values near 130, the critical threshold for P50). It may be unavoidable, but the question of whether it is necessary to add extra errors  by rounding deserves a closer look. In this study we ignore the effect of rounding and analytic errors.

Censoring is more important. For example, the value `$>$9,700' occurs 7 times for site 113, all of them since 2015, whereas the previous maximum reading was 3,683. These high {\em E. coli} counts are important to understanding the underlying state and trend of the river. Changing them to 9,700 would bias downwards the estimates of the mean, the standard deviation, and the trend. The value 9,700 is often only 1 standard deviation above the mean; 12 of the 135 sites have 95th percentile values over 9,700.

We deal with the censored values in the parametric model with a simple, standard approach in which censored values are replaced by their expected values in the model, conditioned on their known ranges. As the parameters now enter into both the model and the data, they obey nonlinear equations. We solve these by an iterative procedure in which the imputed values of the censored data values are iteratively updated. (Of course, the procedure becomes unreliable if too high a proportion of the data is censored.)

In an earlier study \cite{sn}, censored values were removed from the data set before analysing trends. Since they tend to occur later in time, removing them biases trends downwards. The effect can be noticeable. For site 124, in which 3 of 51 values were censored, deleting the censored values gives a trend of $-0.012$ (i.e, 1.2\% improvement per year) and a median {\em E. coli} count in 2020 of 209; replacing censored values with their bounds gives a trend of +0.077 and a median {\em E. coli} count in 2020 of 360; using imputed values gives a trend of +0.103 and a median {\em E. coli} count in 2020 of 421. 

\begin{figure}
\includegraphics[width=46mm]{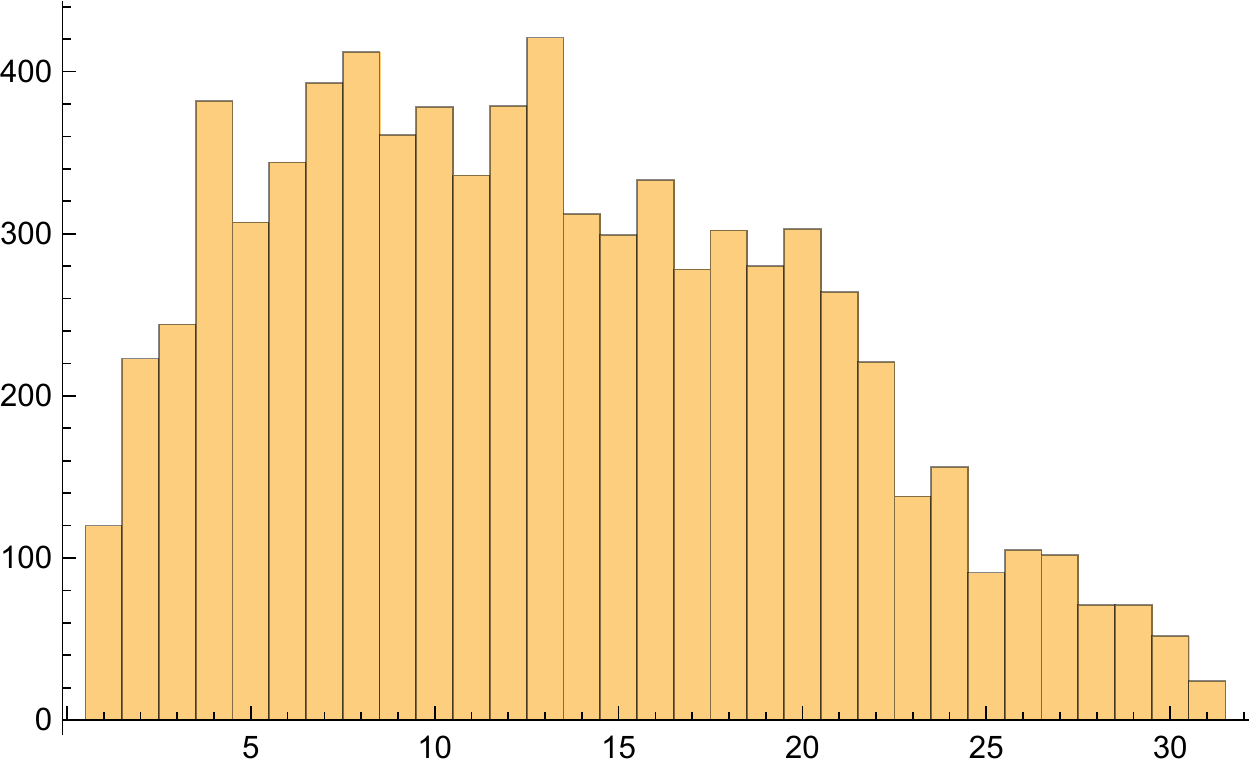}\hspace{5mm}
\includegraphics[width=46mm]{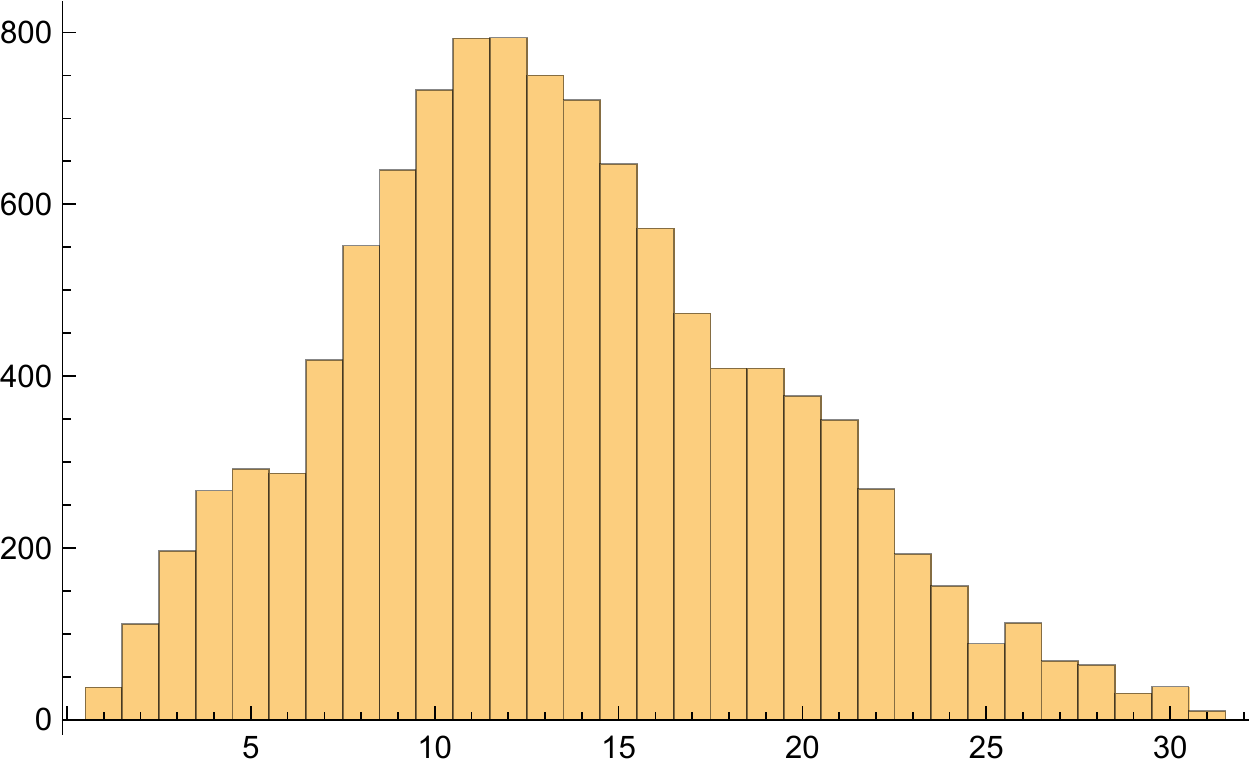}\hspace{5mm}
\small
\begin{overpic}[width=46mm]{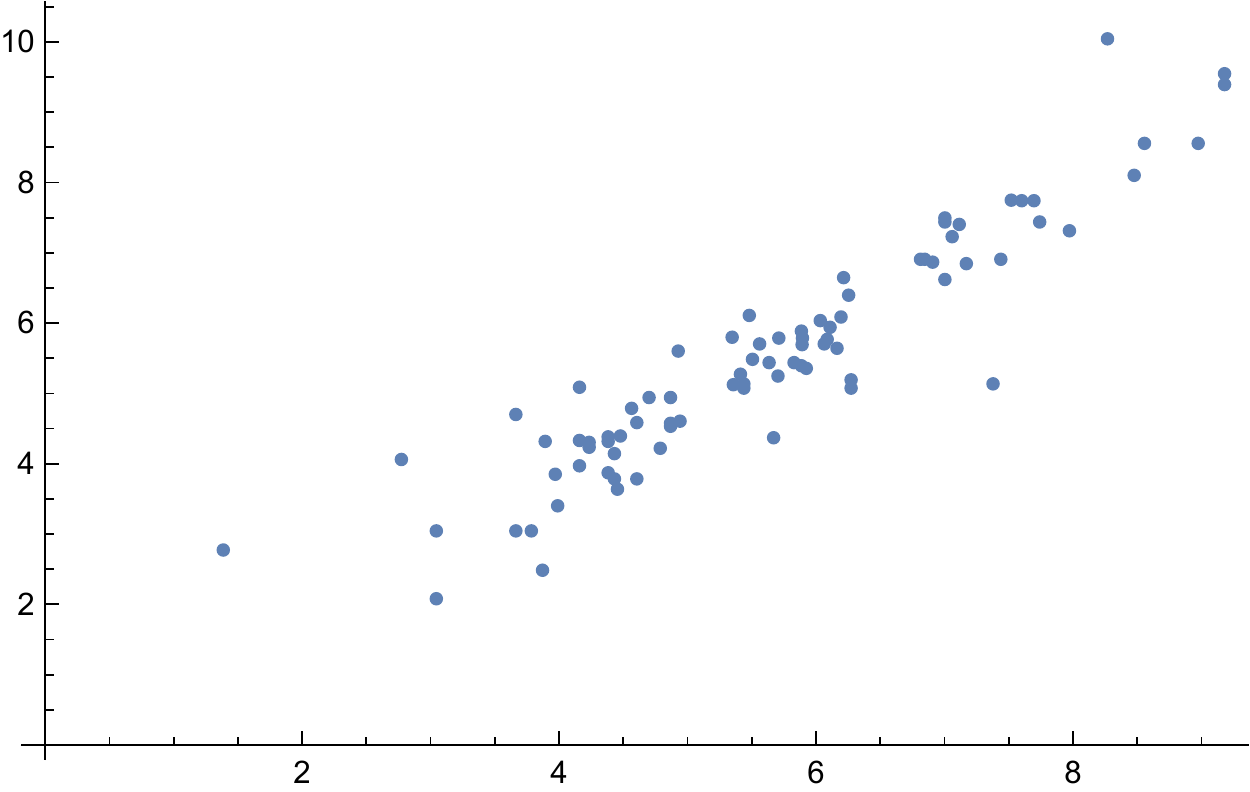}
\put (40,-5) {\tiny Log(E. coli) site 42}
\put (-5, 40) {\tiny \rotatebox{90}{Site 43}}
\end{overpic}
\caption{\label{fig:corr}Histograms of day of the month for data collection. Left: prior to 2013; centre: from 2013.
Right: scatterplot of all same-day readings from nearby sites 42 and 43, showing a correlation coefficient of 0.93.}
\end{figure}

Collecting and analysing the water samples is time-consuming and expensive. A version of the data on LAWA that was available in 2018, but is no longer available, also listed the time of collection, which showed that nearby sites are sampled very nearly at the same time. This makes them correlated. We find that 434 pairs of sites have at least 30 measurements taken on the same day, with correlation between the same-day measurements greater than 0.5. The most highly correlated sites are 42 and 43, two sites on the Manawat\={u} river in Palmerston North, less than 3 km apart. Their correlation coefficient is 0.93 (see Figure~\ref{fig:corr}), which means that the data cannot be pooled to reduce uncertainty in the trend. Sample collection takes place mostly in the middle of the month (Fig.~\ref{fig:corr}), even more so from 2013, which increases the chances of same-day collection, even between distant sites. The correlation may be due to an underlying cause, such as {\em E. coli} outbreaks linked to the same source, or rainfall triggering outbreaks.

\begin{figure}
\begin{center}
\begin{overpic}[width=10cm]{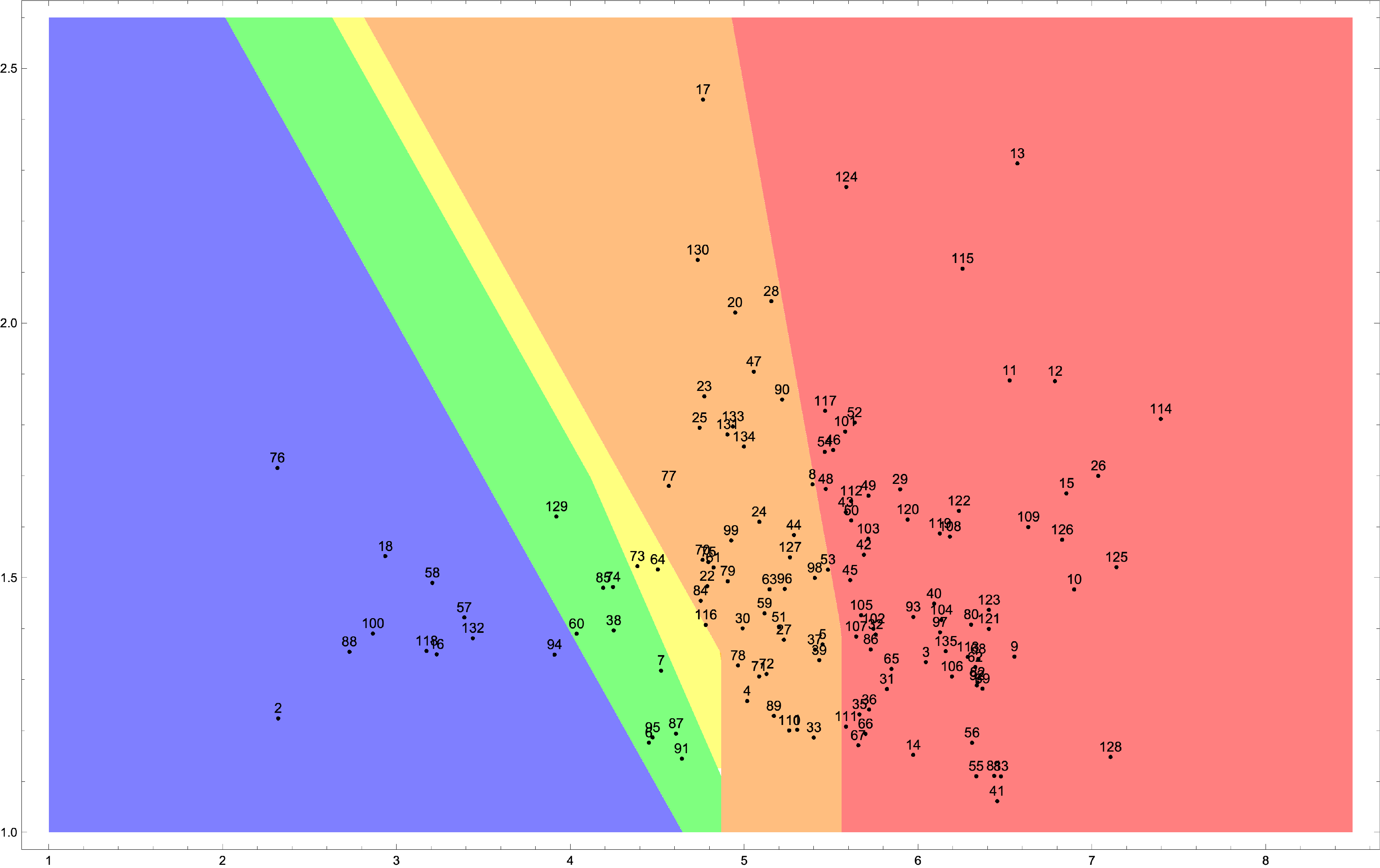}
\put (82,-2){ $\mu$}
\put (-4,56) {$\sigma$}
\end{overpic}\\[10mm]
\caption{\label{fig:jumps}
Results of the parametric model showing freshwater state $(\mu,\sigma)$ estimated using all data for each site. }
\end{center}
\end{figure}

Results for freshwater state using the parametric model are shown in Figure~\ref{fig:jumps} and Tables~\ref{tab:cat} and~\ref{tab:allres}. There are 24 `clean' sites (categories A--C) and 111 `dirty' sites (categories D--E) with no significant systematic difference between the percentile and parametric models.

\section{Trends in the parametric model}
\begin{figure}
\begin{center}
\begin{overpic}[width=8cm]{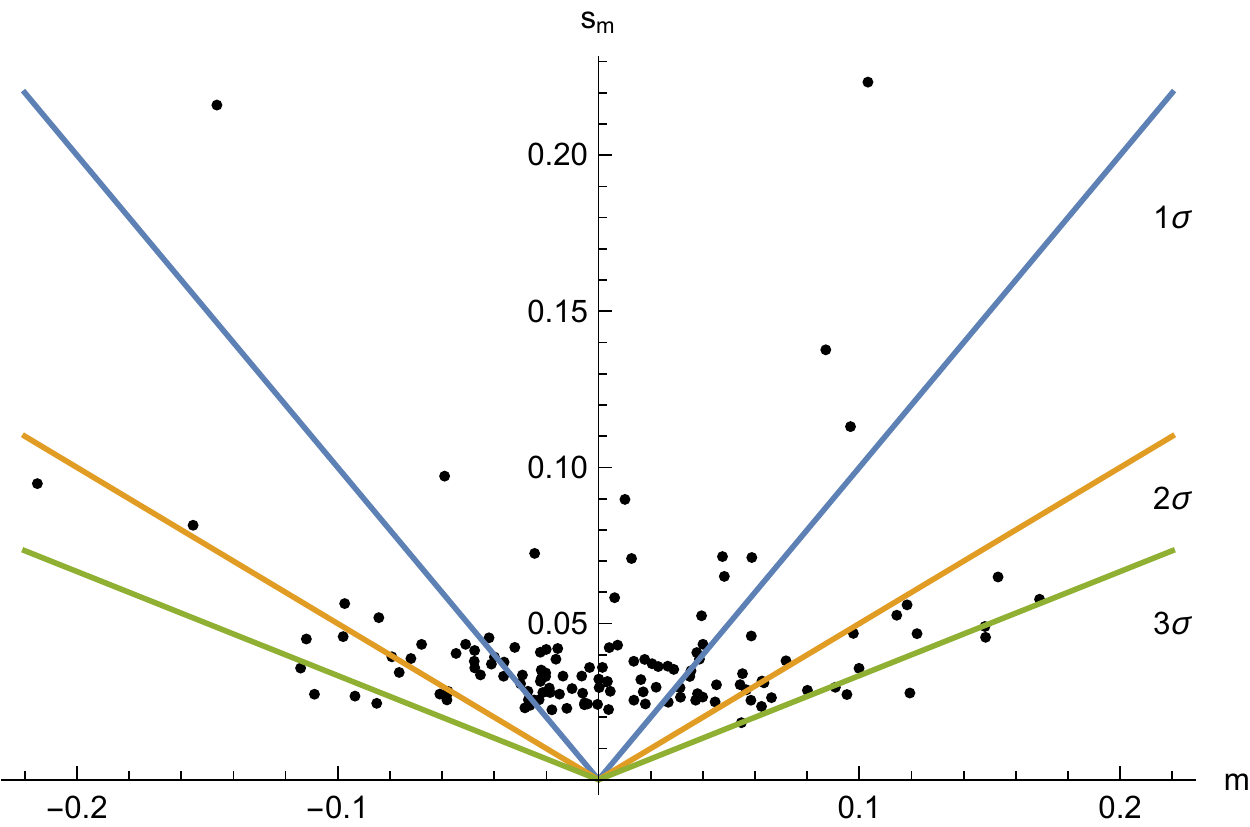}
\put (80,-4){ \small $\mu_1$}
\put (40,70){\tiny standard error}
\end{overpic}
\end{center}
\caption{\label{fig:slopes} Scatterplot of trends for the 135 sites vs. standard error of each trend measurement. 
Regions in which an individual trend is significant at the 1, 2, and 3$\sigma$ level are indicated.
These sampling errors have the effect of smearing out the distribution of measured trends.}
\end{figure}

The lognormal parametric model is convenient for trend determination. We first performed linear regressions to the data $(t_i,L_i)$ for each site. The standard deviations of the residuals in the first half and in the second half of each site were compared. Over the set of all sites, the differences were not significant. Therefore we adopted the model $L_i\sim N(\mu + m t_i,\sigma)$, where the parameter $m$ is the trend, and $\sigma$ is the standard deviation of the residuals. 
%and is smaller than the standard deviation of the data, as used for state determination. 
A value of $m=-0.05$ means a 5\% reduction in {\em E. coli} levels per year.
In this model, the trends in all percentiles are equal, although the  G260, P50, P95, and G540 criteria could be crossed at different times. The measured trends and their standard errors are shown in Figure~\ref{fig:slopes}. 

The state A--E at time $t$ is then determined from the parameters $(\mu+mt,\sigma)$. Note that $\sigma$ is the standard deviation of the residuals and is smaller than the standard deviation of the data, as used for state determination. The standard errors of the parameters are also obtained. We report these at time 2020.

\begin{figure}
\begin{center}
\begin{overpic}[width=12cm]{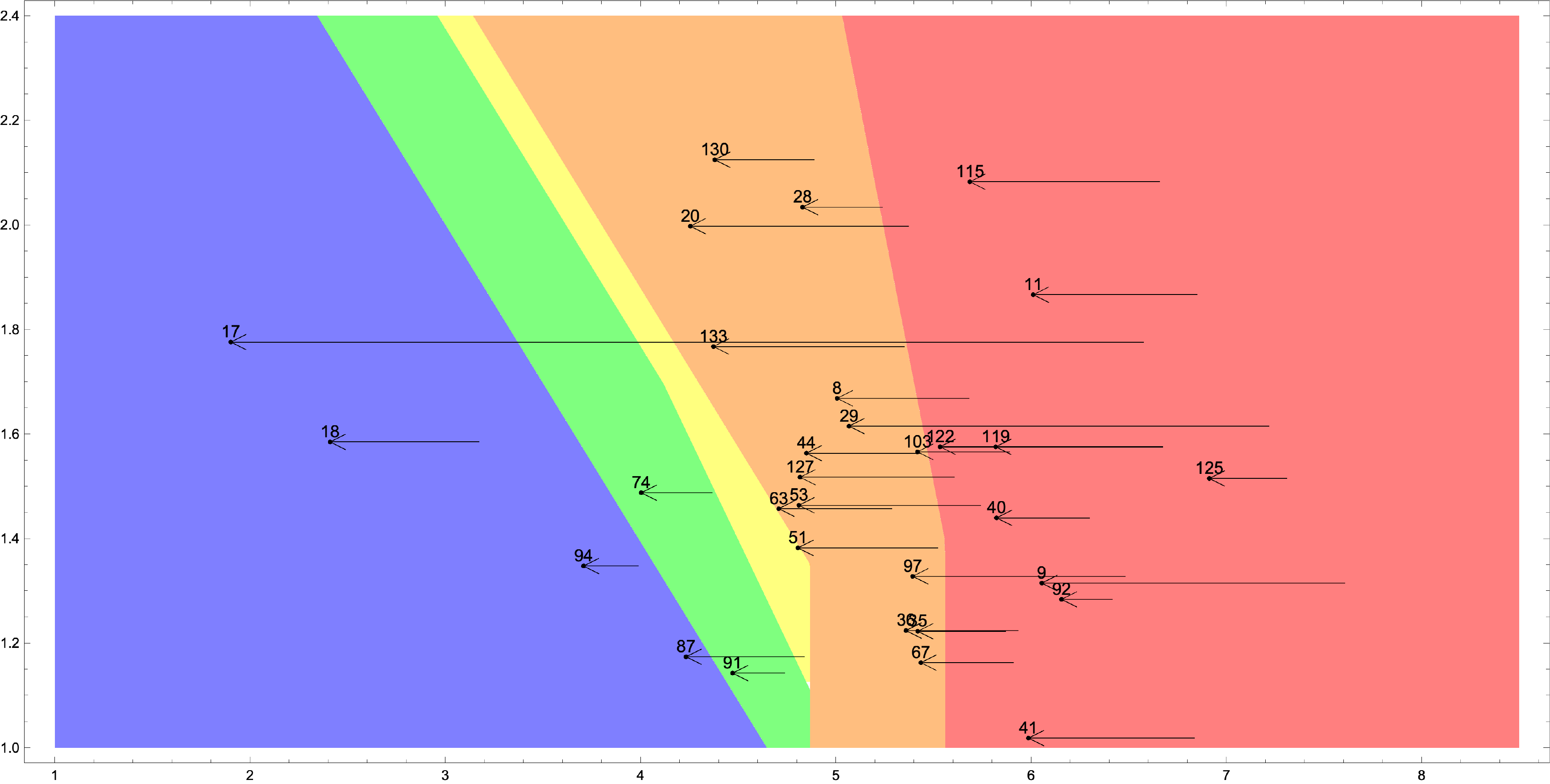}
\put (82,-2){\small $\mu$}
\put (-4,46) {\small $\sigma$}
\end{overpic}\\[10mm]
\begin{overpic}[width=12cm]{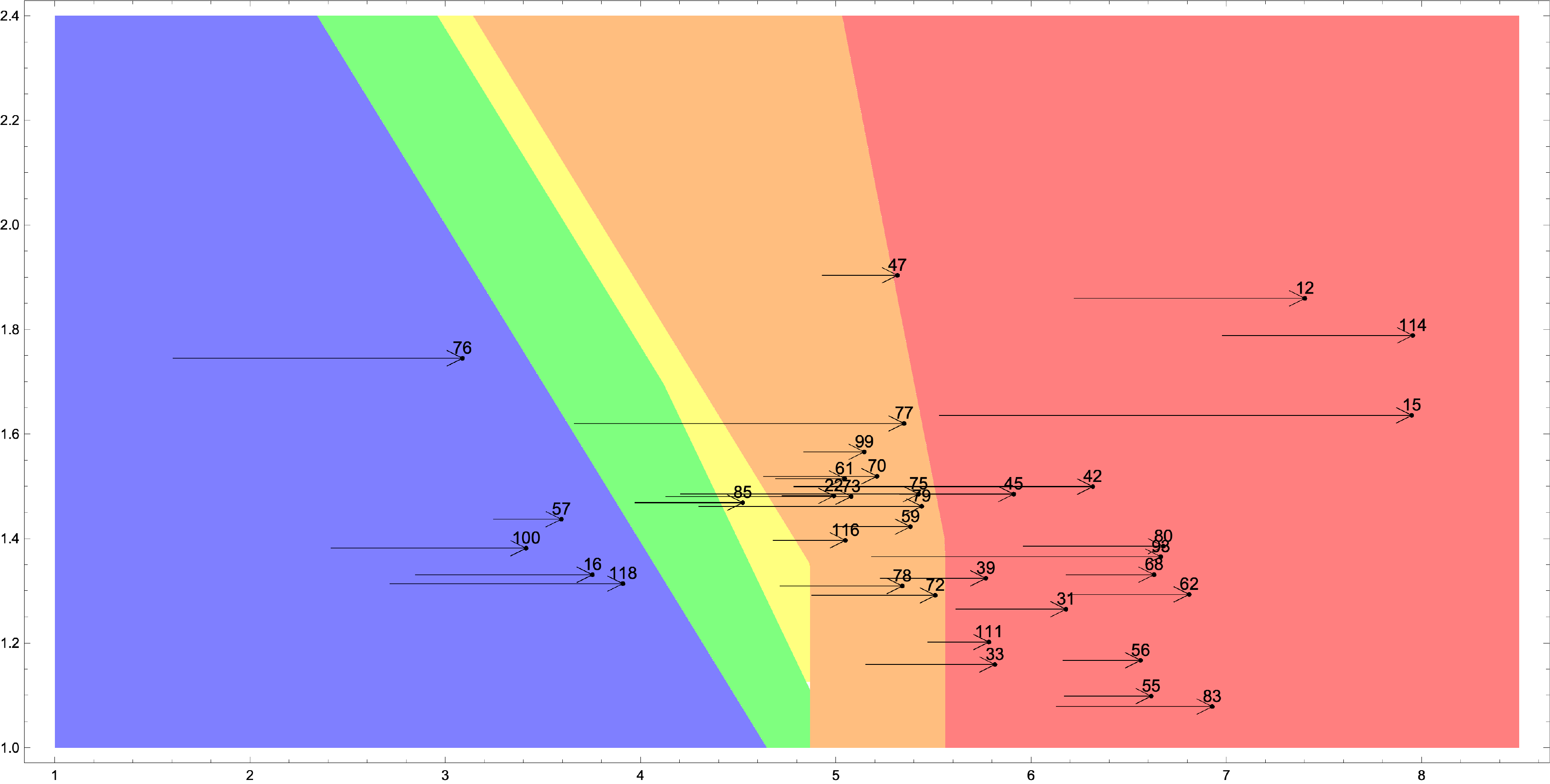}
\put (82,-2){\small $\mu$}
\put (-4,46) {\small$\sigma$}
\end{overpic}\\[10mm]
\begin{overpic}[width=12cm]{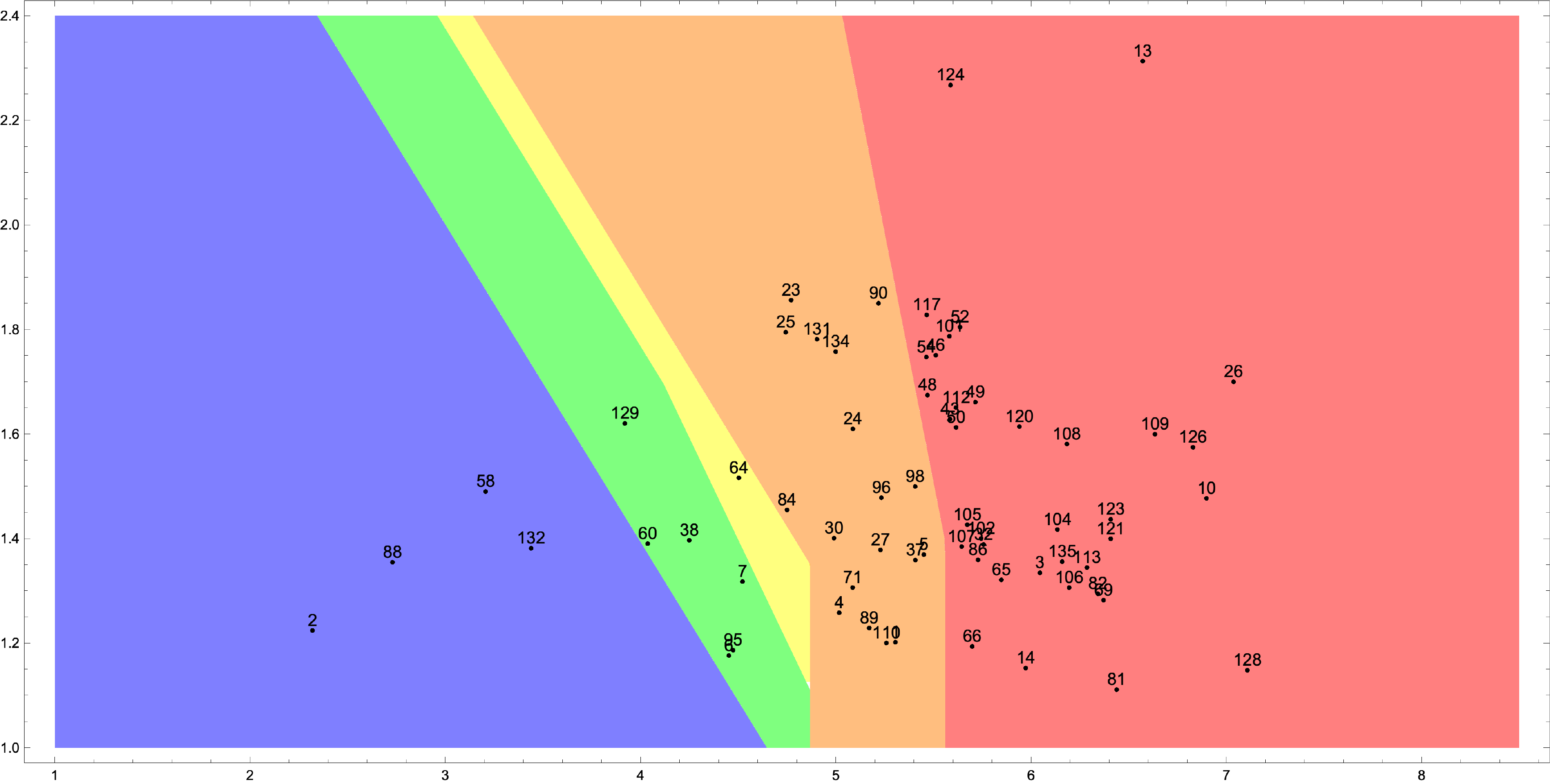}
\put (82,-2){\small$\mu$}
\put (-4,46) {\small$\sigma$}
\end{overpic}\\[10mm]

\caption{\label{fig:jumps2}
Results of the parametric model for trends. Top: Sites where the trend for freshwater quality is improving at the 1$\sigma$ level, with an arrow pointing from the state estimated in 2010 to the state estimated in 2020. Middle: Sites where the trend for freshwater quality is deteriorating at the 1$\sigma$ level. Bottom:
Sites where the trend is not significant at the 1$\sigma$ level, showing the average state.
Note that many of the significant 10-year trends are of the same order as the sizes of the categories.
}
\end{center}
\end{figure}

\begin{figure}
\begin{center}
\includegraphics[width=\textwidth]{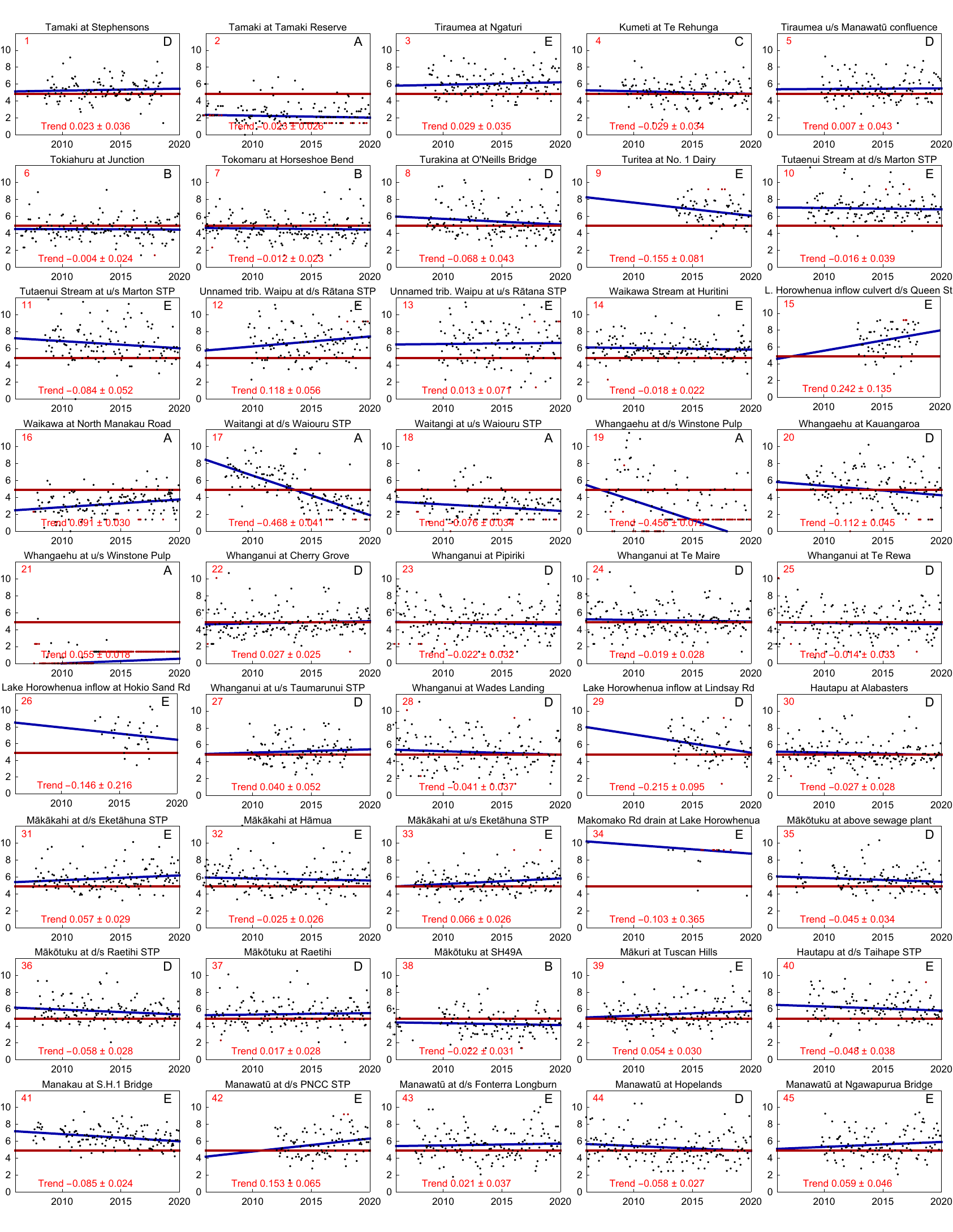}
\end{center}
\caption{\label{trends1}Log {\em E. coli} measurements and trends for sites 1--45. The red and blue lines show the P50 criterion and its trend, respectively. The trend, its standard error, and category in 2020 estimated from the trend are also shown. Censored data (for which imputed values are used) are shown in red.}
\end{figure}

\begin{figure}
\begin{center}
\includegraphics[width=\textwidth]{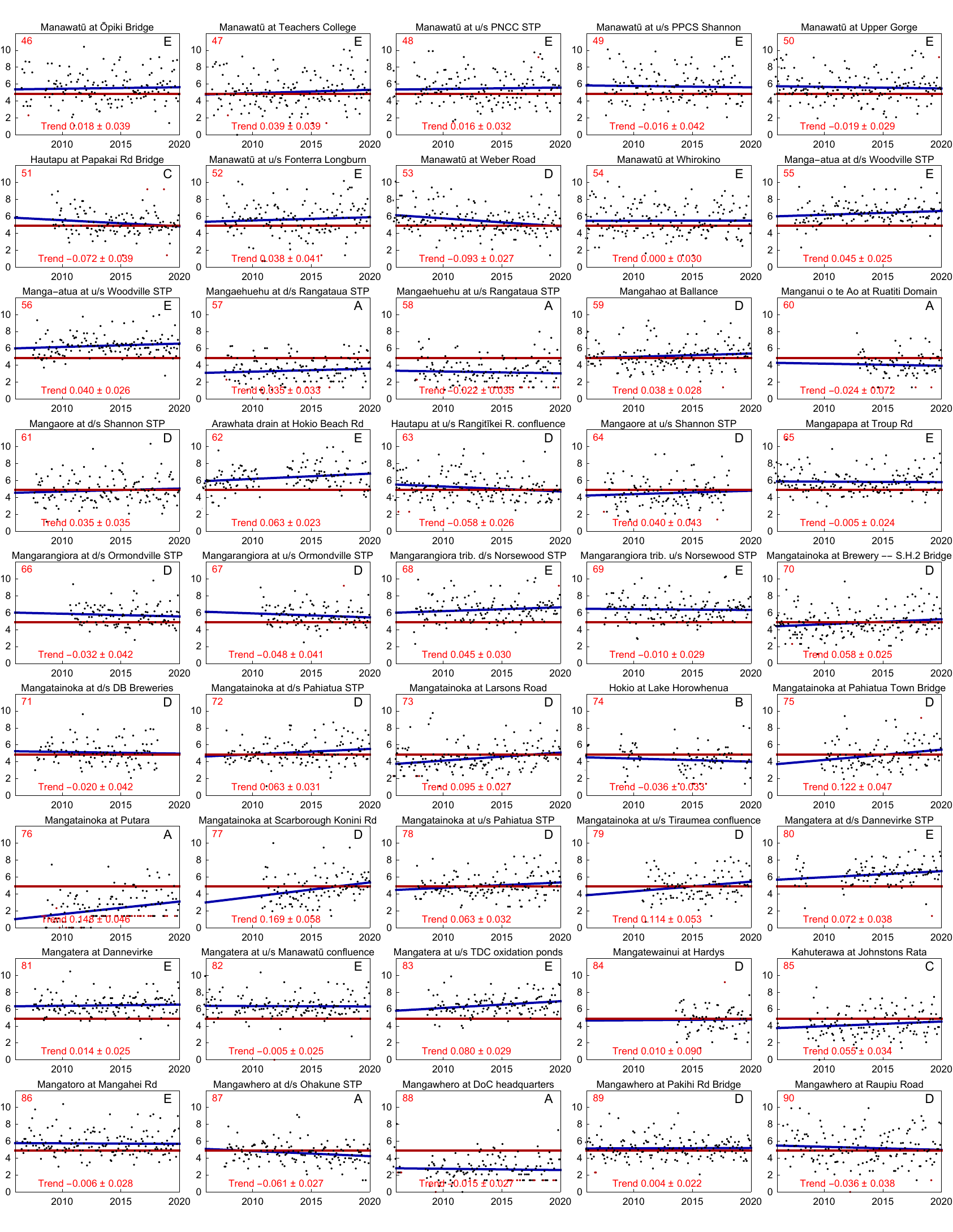}
\end{center}
\caption{\label{trends2}Log {\em E. coli} measurements and trends for sites 46--90. The red and blue lines show the P50 criterion and its trend, respectively. The trend, its standard error, and category in 2020 estimated from the trend are also shown. Censored data (for which imputed values are used) are shown in red.
}
\end{figure}

\begin{figure}
\begin{center}
\includegraphics[width=\textwidth]{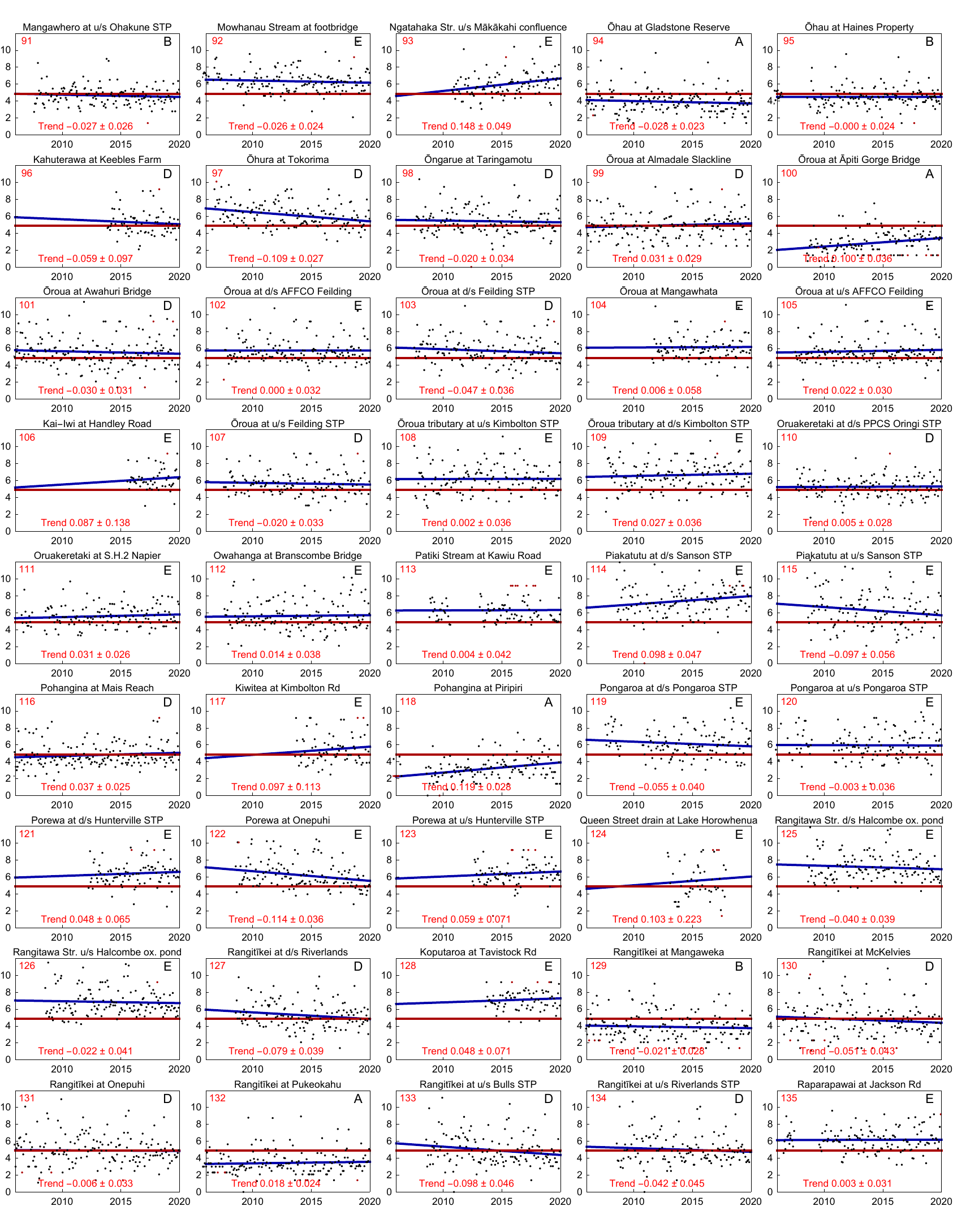}
\end{center}
\caption{\label{trends3}Log {\em E. coli} measurements and trends for sites 91--135. The red and blue lines show the P50 criterion and its trend, respectively. The trend, its standard error, and category in 2020 estimated from the trend are also shown. Censored data (for which imputed values are used) are shown in red.
}
\end{figure}

The results are shown in 
Table~\ref{tab:cat}
 (numbers of sites in each category in 2020); 
Figure~\ref{fig:jumps2} (a visualisation of the impact of 10-year trends on the 2020 category);
Figures~\ref{trends1}--\ref{trends3} (each data set plotted separately along with its trend); and
Table~\ref{tab:allres} (numerical state and trend results for each site).

Of the 135 sites, 17 sites are improving at the 1-sigma level, 9 at least at the 2-sigma level, and 6 at the 3-sigma level. 
18 sites are deteriorating at least at the 1-sigma level, 13 at least at the 2-sigma level, and 5 at the 3-sigma level.
At 67 sites (50\%) the trend was not significant at the 1-sigma level. The average trend is indistinguishable from 0.

However, each measured trend is subject to a sampling error. These have a median value of 0.034. This has the effect of smearing out the distribution of observed trends. (If all true trends were equal, we would observe a normal distribution of trends with standard deviation 0.053.) If the true trends are normally distributed with variance $\sigma_1^2$ and the sampling errors are independent and normally distributed with
mean 0 and variance $\sigma_2^2$, then the observed trends will be normally distributed with variance $\sigma_1^2 + \sigma_2^2$. Under these assumptions, the true distribution of trends is likely to be more concentrated than shown in the figure. However, the correlation between measurements at nearby sites makes it difficult to take this observation further.

Trends that are significant at less than the 1-sigma level should be treated with caution. (Sites with {\em any} improving trend have been called `more likely to be improving than not' \cite{sn}.) Site 47, Manawat\={u}  at Teacher's College, showed a quite strong improving trend of about $-0.15$ (i.e. 15\% per year) with $z_m\approx 1.3$ consistently from early 2013 to mid 2017. This was then reversed by several particularly high readings of up to 23,000. It is now showing a deteriorating trend of $+0.04$ with $z_m=1.0$.
It is not possible to determine from the data whether the  water quality was actually improving until mid-2017 or whether this was just an artefact of noise, as the pattern seen here is typical of independent random samples.

\paragraph{How many samples are needed to detect a trend?}
For data drawn from the $(\mu,\sigma)$ model with $n$ uniformly-spaced samples per year for $T$ years, the expected standard error of the trend is $\sqrt{12}\sigma T^{-3/2} n^{-1/2}$ \cite{ah-fe}.  Adopting the (very minimal) requirement that to detect a trend $m$ requires the expected standard error to be less than $|m|$, we need  $n>\frac{12\sigma^2}{T^3 m^2}.$

Consider a site with a typical value of $\sigma=1.5$. With 10 years of data and 12 samples per year, trends $|m|>0.047$ are detectable; with 52 samples per year, trends $|m|>0.023$ are detectable.

However, with 5 years of data and 12 samples per year, trends $|m|>0.13$ are detectable; with 52 samples per year, trends $|m|>0.064$ are detectable. In our dataset, only 5 of the 135 sites had $m<-0.13$ (sites 9, 17, 19, 26, 29) and some of these are exceptional (e.g., site 17, significant at $z=11$ due to the installation of a new water treatment plant, and site 19 due to highly  acidic water in the Whangaehu river, as shown in site 21 at which scarcely any {\em E. coli} are ever detected.)

Conversely, a number of sites have trends in the range $-0.07$ to $-0.10$---good enough to move from the grade D/E boundary to the C/D boundary in 10 years. To detect this in 5 years even at the 1$\sigma$ level needs weekly monitoring.

We conclude that to get useful trends in 5 years requires more frequent measurements than is the current practice.


\begin{table}
\begin{center}
\begin{tabular}{ l ccccc }
\hline\hline
& A & B & C & D & E\\
\hline
Percentile state & 12 & 8 & 2 & 60 & 53\\
Average state & 12 & 10 & 2 & 44 & 67\\
2020 state & 16 & 7 & 3 & 44 & 61\\ \hline\hline
\end{tabular}
\end{center}
\caption{\label{tab:cat}
Number of Manawat\=u-Whanganui sites in each 
water quality category determined in two ways. Top row: Standard method, based on percentiles. Middle: 
from the parametric model, determining the average state using all data at each site. Bottom: From the parametric model, estimating the state in January 2020 using the fitted trend.}
\end{table}

%\begin{table}
%\caption{Results of the linear parametric model. $N_i$: number of measurements at site $i$; Cat$_1$: water quality category using all data and Hazen percentiles; $(\mu_{s_\mu},\sigma)$: mean (and its standard error) and standard deviation and category Cat$_2$, all data; trend $m$ and $z$-score $z_m$ of the trend; 
%mean (and its standard error), standard deviation, and category Cat$_3$ at 2020 from the linear model. Sites are sorted by 2020 category, and then in order of most improving to most deteriorating. Sites that are improving (resp. deteriorating) by at least 0.05 (i.e., by at least 5\% per year) at the 1$\sigma$ level are marked $\uparrow$ (resp. $\downarrow$).}
%\end{table}
%
%\begin{table}
%{\small\hspace{1mm}
%$i$ \hspace{10mm}
%Name of site $i$ \hspace{37mm}
%$N_i$ \hspace{1mm}
%Cat$_1$ \hspace{1mm}
%$\mu_{s_\mu}$ \hspace{2mm}
%$\sigma$ \hspace{2mm}
%Cat$_2$ \hspace{1mm}
%$m$ \hspace{2mm}
%$z_m$ \hspace{4mm}
%$\mu^{2020}_{s_\mu}$ \hspace{3mm}
%$\sigma^{2020}$ \hspace{1mm}
%Cat$_3$}\\
%\begin{center}
%\includegraphics[width=\textwidth]{table.pdf}
%\end{center}
%\end{table}
%
%\begin{table}
%{\small\hspace{1mm}
%$i$ \hspace{10mm}
%Name of site $i$ \hspace{37mm}
%$N_i$ \hspace{1mm}
%Cat$_1$ \hspace{1mm}
%$\mu_{s_\mu}$ \hspace{2mm}
%$\sigma$ \hspace{2mm}
%Cat$_2$ \hspace{1mm}
%$m$ \hspace{2mm}
%$z_m$ \hspace{4mm}
%$\mu^{2020}_{s_\mu}$ \hspace{3mm}
%$\sigma^{2020}$ \hspace{1mm}
%Cat$_3$}\\
%\begin{center}
%\includegraphics[width=\textwidth]{table2.pdf}
%\end{center}
%\end{table}

\section{Discussion}

\paragraph{Observations}
\begin{enumerate}
\item
Looking at the data in Figure~\ref{fig:jumps}, the classification system based on four different criteria  of two different types (percentiles and percentage exceedances) seems to be unnecessarily complex. Very similar classifications, having a very similar relationship to health risks, can be obtained by a single criterion based on a single percentile. The 67\% percentile is easier to measure reliably than the 95\%, and would classify the data just as well. It would also allow the river state to be summarized by the single number $\mu + z_{0.67}\sigma$, which also retains more information that a classification into five classes.
\item
This is backed up by Hunter \cite{hunter}, who concluded, ``the estimation of the 95th percentile using any of the methods
examined here offers little advantage over the current
percentage exceedence approach, other than offering a false
sense of increased accuracy\dots 
It is concluded that a move to use of 95th percentile
calculations in determining compliance with bathing water
standards has no statistical validity and has a number of
disadvantages.''
\item
The definition of category C means that very few sites fall into category C. Changing the category boundaries so that more nearly equal number of sites fall into categories B and C would have allowed the categorisation system to convey more information.
\item
The system used in LAWA \cite{lawa} for reporting states and trends is not transparent. For example, nearly all sites in the region are reported as showing either an improving or a deteriorating trend. Most likely, many of these are not significant. In the parametric model, only 24\% of the trends are significant at the 2$\sigma$ level, not accounting for multiple testing or correlation between sites.
\item
We have not tested the validity of the lognormal model in detail. 95\% of the residuals have $|z|<2$ (consistent with the normal distribution) but 1.2\% have $z<-2$ and 4\% have $z>2$ (cf. 2.2\% in the normal distribution). The discrepancy in $2<z<3$ could repay further examination.
\end{enumerate}

\newpage
\paragraph{Recommendations}
\begin{enumerate}
\item Examine why the data was rounded and if possible, avoid rounding.
\item Examine why the data was truncated and if possible, avoid truncation (censoring). These processes may be losing useful information for no good reason.
\item If possible, use an {\em E. coli} analysis method sensitive down to 1 bacteria/100 ml.
\item Faster, more reliable detection of trends needs more frequent sampling. It is more cost-effective to sample more frequently from fewer sites, as sampling nearby sites at the same time yields almost no  independent information.
\item Re-examine the criteria used for reporting water quality states and trends. In particular, the 95\% criterion is the deciding factor for swimmability for many sites, and should be examined carefully. 
\end{enumerate}

\paragraph{Acknowledgements} This project arose from a problem presented by Horizons Regional Council (HRC) at the 2018 meeting of the Mathematics in Industry Study Group of New Zealand. We thank HRC (particularly Stacey Binsted, Abby Matthews and Manas Chakraborty) for their advice and the other members of the working group (including Ali Abdul Adheem, Jamas Enright, Anton Good, Christian Offen, Cami Sawyer, Sunchai Tongsuksai, Alex White, and Matt Wilkins) who examined various aspects of freshwater quality monitoring in the Horizons region.

%tlrbourdin@gmail.com, 
%npremara@gmail.com, 

\afterpage{%
\begin{landscape}
\setlength{\LTcapwidth}{1.3\textwidth}
\begin{longtable}{rlrcrrrcrrrrrl} \hline\hline 
\multicolumn{1}{c}{$i$} & Name of site $i$ & \multicolumn{1}{c}{$N_i$} & $\mathrm{Cat}_1$ & \multicolumn{1}{c}{$\mu$} & \multicolumn{1}{c}{$s_\mu$} & \multicolumn{1}{c}{$\sigma$} &
    $\mathrm{Cat}_2$ & \multicolumn{1}{c}{$m$} & \multicolumn{1}{c}{$z_m$} & \multicolumn{1}{r}{$\mu_{2020}$} & \multicolumn{1}{c}{$s_{\mu_{2020}}$} & \multicolumn{1}{r}{$\sigma_{2020}$} & \multicolumn{1}{c}{$\mathrm{Cat}_3$}  \\ \hline  \endhead
\hline\hline \\
\caption{Results of the linear parametric model. See last page for the key to the column headings.}
\endfoot
\hline\hline \\
\caption{Results of the percentile, parametric state, and parametric trend models.
$N_i$: number of measurements at site $i$.\\
Cat$_1$: water quality category using all data and Hazen percentiles.\\
 $(\mu,{s_\mu},\sigma,\mathrm{Cat}_2)$: mean (and its standard error), standard deviation, and category, all data, parametric state model.\\
$(m,z_m,\mu_{2020},s_{\mu_{2020}},\sigma_{2020},\mathrm{Cat}_3)$: trend and its $z$-score, 
mean (and its standard error), standard deviation, and category at 2020 in the linear trend model.\\
Sites that are improving (resp.\ deteriorating) by at least 0.05 (i.e., by at least 5\% per year) at the 1-sigma level are marked $\uparrow$ (resp.\ $\downarrow$). \\
Macrons have been added to names as accurately as possible using the sources available to us. In cases of doubt, the name recorded in the New Zealand Geographic Board {\em Gazetteer} has been used.}
\label{endoftable}
\endlastfoot
\label{tab:allres}
\label{startoftable}
$1$ & Tamaki at Stephensons & $121$ & D & $5.31$ & $0.11$ & $1.20$ & D & $0.02$ & $0.63$ & $5.45$ & $0.26$ & $1.20$ & D  \\
$2$ & Tamaki at Tamaki Reserve & $160$ & A & $2.32$ & $0.10$ & $1.22$ & A & $-0.02$ & $0.89$ & $2.04$ & $0.21$ & $1.34$ & A  \\
$3$ & Tiraumea at Ngaturi & $135$ & E & $6.05$ & $0.11$ & $1.33$ & E & $0.03$ & $0.82$ & $6.21$ & $0.23$ & $1.33$ & E  \\
$4$ & Kumeti at Te Rehunga & $132$ & D & $5.02$ & $0.11$ & $1.26$ & D & $-0.03$ & $0.87$ & $4.86$ & $0.22$ & $1.25$ & C  \\
$5$ & Tiraumea u/s Manawat\=u confluence & $119$ & D & $5.45$ & $0.13$ & $1.37$ & D & $0.01$ & $0.17$ & $5.49$ & $0.26$ & $1.37$ & D  \\
$6$ & Tokiahuru at Junction & $157$ & B & $4.45$ & $0.09$ & $1.18$ & B &$0.00$& $0.17$ & $4.42$ & $0.19$ & $1.18$ & B  \\
$7$ & Tokomaru at Horseshoe Bend & $175$ & B & $4.52$ & $0.10$ & $1.32$ & B & $-0.01$ & $0.53$ & $4.43$ & $0.20$ & $1.32$ & B  \\
$8$ & Turakina at ONeills Bridge & $137$ & D & $5.39$ & $0.14$ & $1.68$ & D & $-0.07$ & $1.57$ & $5.01$ & $0.29$ & $1.67$ & D  $\uparrow$  \\
$9$ & Turitea at No.\ 1 Dairy & $76$ & E & $6.55$ & $0.15$ & $1.35$ & E & $-0.16$ & $1.91$ & $6.05$ & $0.30$ & $1.31$ & E  $\uparrow$  \\
$10$ & Tutaenui Stream at d/s Marton STP & $136$ & E & $6.90$ & $0.13$ & $1.48$ & E & $-0.02$ & $0.42$ & $6.81$ & $0.25$ & $1.48$ & E  \\
$11$ & Tutaenui Stream at u/s Marton STP & $119$ & E & $6.53$ & $0.17$ & $1.89$ & E & $-0.08$ & $1.62$ & $6.01$ & $0.36$ & $1.87$ & E  $\uparrow$  \\
$12$ & Unnamed tributary of Waipu at d/s R\=atana STP & $124$ & E & $6.79$ & $0.17$ & $1.89$ & E & $0.12$ & $2.11$ & $7.40$ & $0.33$ & $1.86$ & E  $\downarrow$  \\
$13$ & Unnamed tributary of Waipu at u/s R\=atana STP & $118$ & E & $6.57$ & $0.21$ & $2.31$ & E & $0.01$ & $0.18$ & $6.64$ & $0.43$ & $2.32$ & E  \\
$14$ & Waikawa Stream at Huritini & $168$ & E & $5.97$ & $0.09$ & $1.15$ & E & $-0.02$ & $0.80$ & $5.86$ & $0.17$ & $1.15$ & E  \\
$15$ & Lake Horowhenua inflow at culvert d/s Queen St & $64$ & E & $6.85$ & $0.21$ & $1.67$ & E & $0.24$ & $1.79$ & $7.95$ & $0.64$ & $1.64$ & E  $\downarrow$  \\
$16$ & Waikawa at North Manakau Road & $154$ & A & $3.23$ & $0.11$ & $1.35$ & A & $0.09$ & $3.07$ & $3.75$ & $0.21$ & $1.33$ & A  $\downarrow$  \\
$17$ & Waitangi at d/s Waiouru STP & $145$ & E & $4.76$ & $0.20$ & $2.44$ & D & $-0.47$ & $11.33$ & $1.90$ & $0.29$ & $1.78$ & A  $\uparrow$  \\
$18$ & Waitangi at u/s Waiouru STP & $141$ & A & $2.94$ & $0.13$ & $1.54$ & A & $-0.08$ & $2.22$ & $2.41$ & $0.25$ & $1.58$ & A  $\uparrow$  \\
$19$ & Whangaehu at d/s Winstone Pulp & $149$ & D & $2.44$ & $0.25$ & $3.08$ & D & $-0.46$ & $6.31$ & $-0.94$ & $0.52$ & $3.16$ & A  $\uparrow$  \\
$20$ & Whangaehu at Kauangaroa & $142$ & D & $4.95$ & $0.17$ & $2.02$ & D & $-0.11$ & $2.49$ & $4.26$ & $0.32$ &$2.00$& D  $\uparrow$  \\
$21$ & Whangaehu at u/s Winstone Pulp & $145$ & A & $0.90$ & $0.07$ & $0.81$ & A & $0.05$ &$3.00$& $0.55$ & $0.13$ & $0.78$ & A  $\downarrow$  \\
$22$ & Whanganui at Cherry Grove & $183$ & D & $4.79$ & $0.11$ & $1.48$ & D & $0.03$ & $1.08$ & $4.99$ & $0.22$ & $1.48$ & D  \\
$23$ & Whanganui at Pipiriki & $177$ & D & $4.77$ & $0.14$ & $1.86$ & D & $-0.02$ & $0.68$ & $4.60$ & $0.28$ & $1.87$ & D  \\
$24$ & Whanganui at Te Maire & $178$ & D & $5.09$ & $0.12$ & $1.61$ & D & $-0.02$ & $0.67$ & $4.95$ & $0.24$ & $1.61$ & D  \\
$25$ & Whanganui at Te Rewa & $167$ & D & $4.74$ & $0.14$ & $1.79$ & D & $-0.01$ & $0.41$ & $4.64$ & $0.28$ & $1.80$ & D  \\
$26$ & Lake Horowhenua inflow at Hokio Sand Rd & $32$ & E & $7.04$ & $0.30$ & $1.70$ & E & $-0.15$ & $0.68$ & $6.46$ & $0.90$ & $1.69$ & E  \\
$27$ & Whanganui at u/s Taumarunui STP & $104$ & D & $5.23$ & $0.14$ & $1.38$ & D & $0.04$ & $0.75$ & $5.47$ & $0.34$ & $1.37$ & D  \\
$28$ & Whanganui at Wades Landing & $160$ & E & $5.16$ & $0.16$ & $2.04$ & D & $-0.04$ & $1.11$ & $4.83$ & $0.33$ & $2.03$ & D  \\
$29$ & Lake Horowhenua inflow at Lindsay Rd & $75$ & E & $5.90$ & $0.19$ & $1.67$ & E & $-0.22$ & $2.27$ & $5.07$ & $0.41$ & $1.61$ & D  $\uparrow$  \\
$30$ & Hautapu at Alabasters & $162$ & D & $4.99$ & $0.11$ & $1.40$ & D & $-0.03$ & $0.96$ & $4.81$ & $0.22$ & $1.40$ & D  \\
$31$ & M\=ak\=akahi at d/s Eket\=ahuna STP & $149$ & E & $5.82$ & $0.10$ & $1.28$ & E & $0.06$ & $1.97$ & $6.18$ & $0.21$ & $1.26$ & E  $\downarrow$  \\
$32$ & M\=ak\=akahi at H\=amua & $172$ & D & $5.76$ & $0.11$ & $1.39$ & E & $-0.02$ & $0.97$ & $5.58$ & $0.22$ & $1.38$ & E  \\
$33$ & M\=ak\=akahi at u/s Eket\=ahuna STP & $150$ & D & $5.40$ & $0.10$ & $1.19$ & D & $0.07$ & $2.52$ & $5.82$ & $0.19$ & $1.16$ & E  $\downarrow$  \\
$34$ & Makomako Rd drain at Lake Horowhenua & $15$ & E & $8.73$ & $0.53$ & $2.04$ & E & $-0.10$ & $0.28$ & $8.75$ & $1.53$ & $2.35$ & E  \\
$35$ & M\=ak\=otuku at above sewage plant & $123$ & D & $5.66$ & $0.11$ & $1.23$ & E & $-0.05$ & $1.35$ & $5.42$ & $0.21$ & $1.22$ & D  \\
$36$ & M\=ak\=otuku at d/s Raetihi STP & $148$ & D & $5.72$ & $0.10$ & $1.24$ & E & $-0.06$ & $2.04$ & $5.36$ & $0.20$ & $1.22$ & D  $\uparrow$  \\
$37$ & M\=ak\=otuku at Raetihi & $156$ & D & $5.41$ & $0.11$ & $1.36$ & D & $0.02$ & $0.61$ & $5.52$ & $0.22$ & $1.36$ & D  \\
$38$ & M\=ak\=otuku at SH49A & $134$ & B & $4.25$ & $0.12$ & $1.40$ & B & $-0.02$ & $0.71$ & $4.11$ & $0.23$ & $1.40$ & B  \\
$39$ & M\=akuri at Tuscan Hills & $148$ & D & $5.43$ & $0.11$ & $1.34$ & D & $0.05$ & $1.78$ & $5.77$ & $0.22$ & $1.32$ & E  $\downarrow$  \\
$40$ & Hautapu at d/s Taihape STP & $130$ & E & $6.09$ & $0.13$ & $1.45$ & E & $-0.05$ & $1.26$ & $5.82$ & $0.25$ & $1.44$ & E  \\
$41$ & Manakau at S.H.1 Bridge & $142$ & E & $6.46$ & $0.09$ & $1.06$ & E & $-0.09$ & $3.48$ & $5.99$ & $0.16$ & $1.02$ & E  $\uparrow$  \\
$42$ & Manawat\=u at d/s PNCC STP & $97$ & E & $5.69$ & $0.16$ & $1.55$ & E & $0.15$ & $2.36$ & $6.32$ & $0.31$ & $1.50$ & E  $\downarrow$  \\
$43$ & Manawat\=u at d/s Fonterra Longburn & $148$ & D & $5.59$ & $0.13$ & $1.63$ & E & $0.02$ & $0.55$ & $5.71$ & $0.27$ & $1.63$ & E  \\
$44$ & Manawat\=u at Hopelands & $178$ & D & $5.29$ & $0.12$ & $1.58$ & D & $-0.06$ & $2.16$ & $4.85$ & $0.23$ & $1.56$ & D  $\uparrow$  \\
$45$ & Manawat\=u at Ngawapurua Bridge & $120$ & E & $5.61$ & $0.14$ & $1.50$ & E & $0.06$ & $1.27$ & $5.91$ & $0.27$ & $1.49$ & E  $\downarrow$  \\
$46$ & Manawat\=u at \=Opiki Bridge & $148$ & D & $5.51$ & $0.14$ & $1.75$ & E & $0.02$ & $0.46$ & $5.62$ & $0.28$ & $1.75$ & E  \\
$47$ & Manawat\=u at Teachers College & $161$ & D & $5.06$ & $0.15$ & $1.90$ & D & $0.04$ &$1.00$& $5.32$ & $0.30$ & $1.90$ & E  \\
$48$ & Manawat\=u at u/s PNCC STP & $161$ & D & $5.47$ & $0.13$ & $1.67$ & E & $0.02$ & $0.51$ & $5.58$ & $0.26$ & $1.67$ & E  \\
$49$ & Manawat\=u at u/s PPCS Shannon & $138$ & E & $5.72$ & $0.14$ & $1.66$ & E & $-0.02$ & $0.37$ & $5.61$ & $0.31$ & $1.66$ & E  \\
$50$ & Manawat\=u at Upper Gorge & $172$ & D & $5.62$ & $0.12$ & $1.61$ & E & $-0.02$ & $0.64$ & $5.48$ & $0.25$ & $1.61$ & E  \\
$51$ & Hautapu at Papakai Rd Bridge & $130$ & D & $5.20$ & $0.12$ & $1.40$ & D & $-0.07$ & $1.85$ & $4.81$ & $0.24$ & $1.38$ & C  $\uparrow$  \\
$52$ & Manawat\=u at u/s Fonterra Longburn & $148$ & E & $5.64$ & $0.15$ & $1.80$ & E & $0.04$ & $0.93$ & $5.87$ & $0.30$ & $1.80$ & E  \\
$53$ & Manawat\=u at Weber Road & $170$ & D & $5.48$ & $0.12$ & $1.52$ & D & $-0.09$ & $3.49$ & $4.81$ & $0.22$ & $1.46$ & D  $\uparrow$  \\
$54$ & Manawat\=u at Whirokino & $182$ & E & $5.46$ & $0.13$ & $1.75$ & E &$0.00$& $0.01$ & $5.47$ & $0.26$ & $1.75$ & E  \\
$55$ & Manga-atua at d/s Woodville STP & $150$ & E & $6.34$ & $0.09$ & $1.11$ & E & $0.04$ & $1.79$ & $6.62$ & $0.18$ & $1.10$ & E  \\
$56$ & Manga-atua at u/s Woodville STP & $150$ & E & $6.31$ & $0.10$ & $1.18$ & E & $0.04$ & $1.51$ & $6.56$ & $0.19$ & $1.17$ & E  \\
$57$ & Mangaehuehu at d/s Rangataua STP & $148$ & A & $3.39$ & $0.12$ & $1.42$ & A & $0.03$ & $1.06$ & $3.59$ & $0.24$ & $1.44$ & A  \\
$58$ & Mangaehuehu at u/s Rangataua STP & $148$ & A & $3.21$ & $0.12$ & $1.49$ & A & $-0.02$ & $0.63$ & $3.05$ & $0.25$ & $1.53$ & A  \\
$59$ & Mangahao at Ballance & $160$ & D & $5.12$ & $0.11$ & $1.43$ & D & $0.04$ & $1.38$ & $5.38$ & $0.22$ & $1.42$ & D  \\
$60$ & Manganui o te Ao at Ruatiti Domain & $86$ & B & $4.04$ & $0.15$ & $1.39$ & B & $-0.02$ & $0.34$ & $3.94$ & $0.30$ & $1.40$ & A  \\
$61$ & Mangaore at d/s Shannon STP & $147$ & D & $4.82$ & $0.13$ & $1.52$ & D & $0.04$ & $1.02$ & $5.05$ & $0.25$ & $1.51$ & D  \\
$62$ & Arawhata drain at Hokio Beach Rd & $146$ & E & $6.33$ & $0.11$ & $1.32$ & E & $0.06$ & $2.66$ & $6.81$ & $0.21$ & $1.29$ & E  $\downarrow$  \\
$63$ & Hautapu at u/s Rangit\={\i}kei River confluence & $177$ & D & $5.15$ & $0.11$ & $1.48$ & D & $-0.06$ & $2.28$ & $4.71$ & $0.22$ & $1.46$ & D  $\uparrow$  \\
$64$ & Mangaore at u/s Shannon STP & $127$ & D & $4.50$ & $0.13$ & $1.52$ & C & $0.04$ & $0.92$ & $4.79$ & $0.34$ & $1.52$ & D  \\
$65$ & Mangapapa at Troup Rd & $173$ & E & $5.85$ & $0.10$ & $1.32$ & E & $-0.01$ & $0.23$ & $5.81$ & $0.20$ & $1.32$ & E  \\
$66$ & Mangarangiora at d/s Ormondville STP & $111$ & E & $5.70$ & $0.11$ & $1.19$ & E & $-0.03$ & $0.76$ & $5.55$ & $0.23$ & $1.19$ & D  \\
$67$ & Mangarangiora at u/s Ormondville STP & $111$ & D & $5.66$ & $0.11$ & $1.17$ & E & $-0.05$ & $1.15$ & $5.44$ & $0.22$ & $1.16$ & D  \\
$68$ & Mangarangiora tributary at d/s Norsewood STP & $146$ & E & $6.35$ & $0.11$ & $1.34$ & E & $0.05$ & $1.49$ & $6.63$ & $0.22$ & $1.33$ & E  \\
$69$ & Mangarangiora tributary at u/s Norsewood STP & $150$ & E & $6.37$ & $0.10$ & $1.28$ & E & $-0.01$ & $0.35$ & $6.31$ & $0.21$ & $1.28$ & E  \\
$70$ & Mangatainoka at Brewery --- S.H.2 Bridge & $189$ & D & $4.76$ & $0.11$ & $1.54$ & D & $0.06$ & $2.30$ & $5.21$ & $0.23$ & $1.52$ & D  $\downarrow$  \\
$71$ & Mangatainoka at d/s DB Breweries & $114$ & D & $5.09$ & $0.12$ & $1.31$ & D & $-0.02$ & $0.48$ & $4.95$ & $0.31$ & $1.30$ & D  \\
$72$ & Mangatainoka at d/s Pahiatua STP & $137$ & D & $5.13$ & $0.11$ & $1.31$ & D & $0.06$ & $2.04$ & $5.51$ & $0.22$ & $1.29$ & D  $\downarrow$  \\
$73$ & Mangatainoka at Larsons Road & $172$ & D & $4.39$ & $0.12$ & $1.52$ & C & $0.10$ & $3.49$ & $5.08$ & $0.23$ & $1.48$ & D  $\downarrow$  \\
$74$ & Hokio at Lake Horowhenua & $107$ & B & $4.25$ & $0.14$ & $1.48$ & B & $-0.04$ & $1.10$ &$4.00$& $0.26$ & $1.49$ & B  \\
$75$ & Mangatainoka at Pahiatua Town Bridge & $120$ & D & $4.79$ & $0.14$ & $1.53$ & D & $0.12$ & $2.61$ & $5.42$ & $0.28$ & $1.49$ & D  $\downarrow$  \\
$76$ & Mangatainoka at Putara & $136$ & A & $2.32$ & $0.15$ & $1.72$ & A & $0.15$ & $3.25$ & $3.09$ & $0.31$ & $1.74$ & A  $\downarrow$  \\
$77$ & Mangatainoka at Scarborough Konini Rd & $111$ & D & $4.57$ & $0.16$ & $1.68$ & D & $0.17$ & $2.93$ & $5.35$ & $0.31$ & $1.62$ & D  $\downarrow$  \\
$78$ & Mangatainoka at u/s Pahiatua STP & $139$ & D & $4.97$ & $0.11$ & $1.33$ & D & $0.06$ & $1.98$ & $5.34$ & $0.22$ & $1.31$ & D  $\downarrow$  \\
$79$ & Mangatainoka at u/s Tiraumea confluence & $110$ & D & $4.91$ & $0.14$ & $1.49$ & D & $0.11$ & $2.17$ & $5.44$ & $0.28$ & $1.46$ & D  $\downarrow$  \\
$80$ & Mangatera at d/s Dannevirke STP & $115$ & E & $6.31$ & $0.13$ & $1.41$ & E & $0.07$ & $1.89$ & $6.68$ & $0.24$ & $1.39$ & E  $\downarrow$  \\
$81$ & Mangatera at Dannevirke & $145$ & E & $6.44$ & $0.09$ & $1.11$ & E & $0.01$ & $0.53$ & $6.52$ & $0.18$ & $1.11$ & E  \\
$82$ & Mangatera at u/s Manawat\=u confluence & $158$ & E & $6.34$ & $0.10$ & $1.29$ & E & $-0.01$ & $0.22$ & $6.30$ & $0.21$ & $1.29$ & E  \\
$83$ & Mangatera at u/s TDC oxidation ponds & $134$ & E & $6.48$ & $0.10$ & $1.11$ & E & $0.08$ & $2.80$ & $6.93$ & $0.19$ & $1.08$ & E  $\downarrow$  \\
$84$ & Mangatewainui at Hardys & $77$ & C & $4.75$ & $0.17$ & $1.45$ & D & $0.01$ & $0.11$ & $4.78$ & $0.33$ & $1.45$ & D  \\
$85$ & Kahuterawa at Johnstons Rata & $141$ & C & $4.19$ & $0.12$ & $1.48$ & B & $0.06$ & $1.63$ & $4.52$ & $0.24$ & $1.47$ & C  $\downarrow$  \\
$86$ & Mangatoro at Mangahei Rd & $159$ & D & $5.73$ & $0.11$ & $1.36$ & E & $-0.01$ & $0.22$ & $5.69$ & $0.21$ & $1.36$ & E  \\
$87$ & Mangawhero at d/s Ohakune STP & $144$ & B & $4.61$ & $0.10$ & $1.19$ & B & $-0.06$ & $2.22$ & $4.23$ & $0.20$ & $1.17$ & A  $\uparrow$  \\
$88$ & Mangawhero at DoC headquarters & $153$ & A & $2.73$ & $0.11$ & $1.35$ & A & $-0.01$ & $0.55$ & $2.59$ & $0.22$ & $1.40$ & A  \\
$89$ & Mangawhero at Pakihi Rd Bridge & $171$ & D & $5.17$ & $0.09$ & $1.23$ & D &$0.00$& $0.17$ & $5.20$ & $0.19$ & $1.23$ & D  \\
$90$ & Mangawhero at Raupiu Road & $159$ & D & $5.22$ & $0.15$ & $1.85$ & D & $-0.04$ & $0.96$ & $4.98$ & $0.29$ & $1.85$ & D  \\
$91$ & Mangawhero at u/s Ohakune STP & $148$ & B & $4.64$ & $0.09$ & $1.14$ & B & $-0.03$ & $1.05$ & $4.47$ & $0.19$ & $1.14$ & B  \\
$92$ & Mowhanau Stream at footbridge & $162$ & E & $6.34$ & $0.10$ & $1.29$ & E & $-0.03$ & $1.10$ & $6.16$ & $0.19$ & $1.28$ & E  \\
$93$ & Ngatahaka Stream u/s M\=ak\=akahi confluence & $110$ & E & $5.97$ & $0.14$ & $1.42$ & E & $0.15$ & $3.02$ & $6.66$ & $0.26$ & $1.37$ & E  $\downarrow$  \\
$94$ & \=Ohau at Gladstone Reserve & $181$ & A & $3.91$ & $0.10$ & $1.35$ & A & $-0.03$ & $1.23$ & $3.71$ & $0.19$ & $1.35$ & A  \\
$95$ & \=Ohau at Haines Property & $156$ & B & $4.47$ & $0.09$ & $1.19$ & B &$0.00$& $0.02$ & $4.47$ & $0.18$ & $1.19$ & B  \\
$96$ & Kahuterawa at Keebles Farm & $72$ & D & $5.23$ & $0.17$ & $1.48$ & D & $-0.06$ & $0.61$ & $5.05$ & $0.35$ & $1.47$ & D  \\
$97$ & \=Ohura at Tokorima & $158$ & E & $6.13$ & $0.11$ & $1.39$ & E & $-0.11$ & $3.98$ & $5.39$ & $0.21$ & $1.33$ & D  $\uparrow$  \\
$98$ & \=Ongarue at Taringamotu & $146$ & D & $5.41$ & $0.12$ & $1.50$ & D & $-0.02$ & $0.59$ & $5.28$ & $0.25$ & $1.50$ & D  \\
$99$ & \=Oroua at Almadale Slackline & $169$ & D & $4.93$ & $0.12$ & $1.57$ & D & $0.03$ & $1.06$ & $5.15$ & $0.24$ & $1.57$ & D  \\
$100$ & \=Oroua at \=Apiti Gorge Bridge & $137$ & A & $2.86$ & $0.12$ & $1.39$ & A & $0.10$ & $2.80$ & $3.41$ & $0.24$ & $1.38$ & A  $\downarrow$  \\
$101$ & \=Oroua at Awahuri Bridge & $179$ & D & $5.58$ & $0.13$ & $1.79$ & E & $-0.03$ & $0.97$ & $5.36$ & $0.27$ & $1.78$ & D  \\
$102$ & \=Oroua at d/s AFFCO Feilding & $147$ & D & $5.74$ & $0.12$ & $1.40$ & E &$0.00$&$0.00$& $5.74$ & $0.23$ & $1.40$ & E  \\
$103$ & \=Oroua at d/s Feilding STP & $148$ & E & $5.71$ & $0.13$ & $1.58$ & E & $-0.05$ & $1.33$ & $5.42$ & $0.26$ & $1.57$ & D  \\
$104$ & \=Oroua at Mangawhata & $101$ & E & $6.13$ & $0.14$ & $1.42$ & E & $0.01$ & $0.11$ & $6.16$ & $0.28$ & $1.42$ & E  \\
$105$ & \=Oroua at u/s AFFCO Feilding & $153$ & D & $5.67$ & $0.12$ & $1.43$ & E & $0.02$ & $0.75$ & $5.82$ & $0.22$ & $1.42$ & E  \\
$106$ & Kai-Iwi at Handley Road & $54$ & E & $6.20$ & $0.18$ & $1.31$ & E & $0.09$ & $0.63$ & $6.39$ & $0.36$ & $1.30$ & E  \\
$107$ & \=Oroua at u/s Feilding STP & $140$ & D & $5.64$ & $0.12$ & $1.38$ & E & $-0.02$ & $0.61$ & $5.52$ & $0.24$ & $1.38$ & D  \\
$108$ & \=Oroua tributary at u/s Kimbolton STP & $149$ & E & $6.18$ & $0.13$ & $1.58$ & E &$0.00$& $0.04$ & $6.19$ & $0.26$ & $1.58$ & E  \\
$109$ & \=Oroua tributary at d/s Kimbolton STP & $149$ & E & $6.63$ & $0.13$ & $1.60$ & E & $0.03$ & $0.73$ & $6.80$ & $0.26$ & $1.60$ & E  \\
$110$ & Oruakeretaki at d/s PPCS Oringi STP & $140$ & D & $5.26$ & $0.10$ & $1.20$ & D &$0.00$& $0.16$ & $5.29$ & $0.20$ & $1.20$ & D  \\
$111$ & Oruakeretaki at S.H.2 Napier & $148$ & D & $5.59$ & $0.10$ & $1.21$ & E & $0.03$ & $1.19$ & $5.78$ & $0.19$ & $1.20$ & E  \\
$112$ & Owahanga at Branscombe Bridge & $147$ & D & $5.62$ & $0.14$ & $1.65$ & E & $0.01$ & $0.36$ & $5.70$ & $0.27$ & $1.65$ & E  \\
$113$ & Patiki Stream at Kawiu Road & $96$ & E & $6.29$ & $0.14$ & $1.34$ & E &$0.00$& $0.10$ & $6.31$ & $0.26$ & $1.35$ & E  \\
$114$ & Piakatutu at d/s Sanson STP & $133$ & E & $7.40$ & $0.16$ & $1.81$ & E & $0.10$ & $2.09$ & $7.96$ & $0.31$ & $1.79$ & E  $\downarrow$  \\
$115$ & Piakatutu at u/s Sanson STP & $129$ & E & $6.26$ & $0.19$ & $2.11$ & E & $-0.10$ & $1.73$ & $5.69$ & $0.38$ & $2.08$ & E  $\uparrow$  \\
$116$ & Pohangina at Mais Reach & $172$ & D & $4.78$ & $0.11$ & $1.41$ & D & $0.04$ & $1.46$ & $5.05$ & $0.21$ & $1.40$ & D  \\
$117$ & Kiwitea at Kimbolton Rd & $76$ & D & $5.47$ & $0.21$ & $1.83$ & E & $0.10$ & $0.85$ & $5.78$ & $0.42$ & $1.82$ & E  \\
$118$ & Pohangina at Piripiri & $148$ & A & $3.17$ & $0.11$ & $1.36$ & A & $0.12$ & $4.30$ & $3.91$ & $0.21$ & $1.31$ & A  $\downarrow$  \\
$119$ & Pongaroa at d/s Pongaroa STP & $127$ & E & $6.13$ & $0.14$ & $1.59$ & E & $-0.05$ & $1.35$ & $5.82$ & $0.27$ & $1.58$ & E  $\uparrow$  \\
$120$ & Pongaroa at u/s Pongaroa STP & $135$ & E & $5.94$ & $0.14$ & $1.61$ & E &$0.00$& $0.10$ & $5.92$ & $0.26$ & $1.61$ & E  \\
$121$ & Porewa at d/s Hunterville STP & $93$ & E & $6.41$ & $0.15$ & $1.40$ & E & $0.05$ & $0.74$ & $6.60$ & $0.29$ & $1.40$ & E  \\
$122$ & Porewa at Onepuhi & $142$ & E & $6.24$ & $0.14$ & $1.63$ & E & $-0.11$ & $3.20$ & $5.53$ & $0.26$ & $1.58$ & E  $\uparrow$  \\
$123$ & Porewa at u/s Hunterville STP & $87$ & E & $6.41$ & $0.15$ & $1.44$ & E & $0.06$ & $0.83$ & $6.64$ & $0.31$ & $1.43$ & E  \\
$124$ & Queen Street drain at Lake Horowhenua & $51$ & E & $5.59$ & $0.32$ & $2.27$ & E & $0.10$ & $0.46$ & $6.04$ & $1.04$ & $2.25$ & E  \\
$125$ & Rangitawa Stream at d/s Halcombe ox.\ pond & $137$ & E & $7.14$ & $0.13$ & $1.52$ & E & $-0.04$ & $1.02$ & $6.91$ & $0.26$ & $1.52$ & E  \\
$126$ & Rangitawa Stream at u/s Halcombe ox.\ pond & $137$ & E & $6.83$ & $0.13$ & $1.57$ & E & $-0.02$ & $0.55$ & $6.70$ & $0.27$ & $1.57$ & E  \\
$127$ & Rangit\={\i}kei at d/s Riverlands & $136$ & D & $5.26$ & $0.13$ & $1.54$ & D & $-0.08$ & $2.01$ & $4.82$ & $0.26$ & $1.52$ & D  $\uparrow$  \\
$128$ & Koputaroa at Tavistock Rd & $77$ & E & $7.11$ & $0.13$ & $1.15$ & E & $0.05$ & $0.67$ & $7.26$ & $0.26$ & $1.15$ & E  \\
$129$ & Rangit\={\i}kei at Mangaweka & $182$ & D & $3.92$ & $0.12$ & $1.62$ & B & $-0.02$ & $0.76$ & $3.75$ & $0.24$ & $1.64$ & B  \\
$130$ & Rangit\={\i}kei at McKelvies & $160$ & D & $4.73$ & $0.17$ & $2.12$ & D & $-0.05$ & $1.18$ & $4.38$ & $0.34$ & $2.12$ & D  $\uparrow$  \\
$131$ & Rangit\={\i}kei at Onepuhi & $169$ & D & $4.90$ & $0.14$ & $1.78$ & D & $-0.01$ & $0.19$ & $4.85$ & $0.28$ & $1.79$ & D  \\
$132$ & Rangit\={\i}kei at Pukeokahu & $179$ & A & $3.44$ & $0.10$ & $1.38$ & A & $0.02$ & $0.74$ & $3.56$ & $0.21$ & $1.40$ & A  \\
$133$ & Rangit\={\i}kei at u/s Bulls STP & $137$ & D & $4.94$ & $0.15$ & $1.80$ & D & $-0.10$ & $2.14$ & $4.37$ & $0.30$ & $1.77$ & D  $\uparrow$  \\
$134$ & Rangit\={\i}kei at u/s Riverlands STP & $137$ & D &$5.00$& $0.15$ & $1.76$ & D & $-0.04$ & $0.92$ & $4.76$ & $0.30$ & $1.75$ & D  \\
$135$ & Raparapawai at Jackson Rd & $137$ & E & $6.16$ & $0.12$ & $1.36$ & E &$0.00$& $0.11$ & $6.18$ & $0.22$ & $1.36$ & E \\
\end{longtable}
\end{landscape}}

\end{document}